\date{}
\documentclass{article}

\usepackage[linesnumbered,ruled]{algorithm2e}
\usepackage{amsmath} 
\usepackage[a4paper]{geometry}
\usepackage{apacite}
\usepackage{natbib}
\usepackage{float}
\usepackage{graphicx} 
\usepackage{latexsym,amsmath,amsfonts,amsthm,amssymb,graphicx,color,mathrsfs}
\usepackage{xcolor}
\usepackage{subfigure}
\usepackage{subcaption}
\usepackage{tabularx}
\usepackage{booktabs}
\usepackage{bm}
\usepackage{tikz}
\usepackage{multirow} 
\usepackage{adjustbox}
\usepackage{array} 
\usepackage{listings}%
\numberwithin{equation}{section}
\theoremstyle{plain}

\theoremstyle{plain}

\theoremstyle{definition}

\usepackage{caption}
\usepackage[nottoc]{tocbibind}
\allowdisplaybreaks[4]
\usepackage{dsfont}
\DeclareMathAlphabet{\mathcal}{OMS}{cmsy}{m}{n}

\providecommand{\lemmaname}{Lemma}

\providecommand{\theoremname}{Theorem}

\makeatother

\providecommand{\definitionname}{Definition}
\providecommand{\lemmaname}{Lemma}
\providecommand{\theoremname}{Theorem}

\usepackage{authblk}

\begin{document}
\title{Signature Decomposition Method Applying to Pair Trading}


\newcommand{\corrauth}[1]{#1$^*$}
\newcommand{\equalcont}[1]{#1$^\dagger$}

\author[1]{\equalcont{Zihao Guo}}
\author[2,3]{\equalcont{Hanqing Jin}}
\author[2]{\corrauth{Jiaqi Kuang$^\dagger$}}
\author[2,3]{\equalcont{Zhongmin Qian}}
\author[1]{\equalcont{Jinghan Wang}}

\affil[1]{Zhongtai Securities Institute for Financial Studies, Shandong University, No. 27 Shanda South Road, Jinan, 250100, Shandong, China}
\affil[2]{Mathematical Modelling and Data Analytics Center, Oxford Suzhou Centre For Advanced Research, No. 388 Ruoshui Road, Suzhou, 215000, Jiangsu, China}
\affil[3]{Mathematical Institute, University of Oxford, Andrew Wiles Building, Radcliffe Observatory Quarter, Oxford, OX2 6GG, Oxfordshire, United Kingdom}



\maketitle
\begin{abstract}
    High-frequency quantitative trading strategies have long been of significant interest in futures market. While advanced statistical arbitrage and deep learning enhance high-frequency data processing, they diminish opportunities for traditional methods and yield less interpretable, unstable strategies. Consequently, developing stable, interpretable quantitative strategies remains a priority in futures markets. In this study, we propose a novel pair trading strategy by leveraging the mathematical concept of path signature which serves as a feature representation of time series. Specifically, the path signature is decomposed into two new indicators: the path interactivity indicator segmented signature and the directional indicator covariation of increments, which serve as double filters in strategy design. Empirical experiments using minute-level futures data show our strategy significantly outperforms traditional pair trading, delivering higher returns, lower maximum drawdown, and higher Sharpe ratio. The proposed method enhances interpretability and robustness while maintaining strong returns, demonstrating the potential of path signatures in financial trading.

\medskip{}
\end{abstract}

\newpage

\section{Introduction}
In financial market, robust, consistently profitable strategies and indicators are pursued by participators. Among them, arbitrage strategies are generally considered as a comparatively lower-risk investment approach, hedging most market risks through simultaneously buying and selling multiple financial assets \citep{dybvig1989arbitrage, yadav1990stock, liew2013pairs,  krauss2017statistical}. Pair trading \citep{vidyamurthy2004pairs, elliott2005pairs} is one of the popular and widely used statistical arbitrages in trading stocks \citep{chen2019empirical}, futures and options \citep{draper2002study}. The core concept of pair trading involves trading the price spread between two correlated assets. When their prices spread deviates from the expected value, the pair trading strategy involves going long on one relatively undervalued asset while shorting the other relatively overvalued one and closing the positions for profit after the spread reverts to normal levels. Arbitrage opportunities typically arise in imperfect markets. However, as more market participants exploit these opportunities, the price spread tends to disappear quickly \citep{krauss2017statistical}. What's more, in highly complex financial markets, just capturing linear correlations alone is insufficient \citep{elliott2005pairs}, as the price spread is often not merely a simple price difference or ratio, but rather a nonlinear function composed of multiple features. So, to better capture arbitrage opportunities, an interpretable pair trading strategy that comprehensively incorporates both linear and nonlinear structures is of significant interest and importance.

Extracting nonlinear features and capturing complex patterns from financial time series data is always one of the main tasks in financial data analysis. It is generally believed that models capable of capturing nonlinear features in time series data should be more effective than traditional linear models. The underlying reason is that flows of financial data typically exhibit high levels of nonlinearity, random, and complex structures, which are usually difficult to be fully captured with linear models. Some studies have explored various nonlinear approaches to enhance statistical arbitrage strategies. For instance, the Hurst exponent has been applied to identify mean-reverting regimes in equity pairs, improving timing accuracy in pair trades \citep{bui2022applying}. Graph-based clustering algorithms have also been used to construct multi-pair portfolios by detecting latent correlation structures in large asset universes \citep{korniejczuk2024statistical}. These studies provide valuable insights for nonlinear methods, but may not be effective in capturing short-term, path-dependent interactions that are crucial in high-frequency trading. Recently, with development of artificial intelligence technology, an increasing number of machine learning and deep learning models with advanced capability have been applied to extract nonlinear features from financial time series data, such as Multi-Layer Perceptron (MLP) \citep{Jasemi2011,enkhbayar2025predictive} for extracting nonlinear signal, Support Vector Machine (SVM) \citep{li2014, barboza2017} for selection of nonlinear characteristic, Deep Neural Network (DNN) \citep{songyoo2019} for capture nonlinear relationships, Long Short-Term Memory (LSTM) \citep{yuyan2020, kimcho2022, phuoc2024, kashif2025lstm} and Transformer \citep{stevenson2021} for nonlinear analysis and decisions of financial time series. However, a known issue with machine and deep learning models applied in financial strategies is their poor interpretability. Lacking of sufficient theoretical foundation, the training process operates as a black box, offering limited controllability, which might result in unstable and confused decisions. As a consequence, applications of these models currently used in the financial industry, particularly in the field of quantitative strategies, need to be approached with caution in practice.

Therefore, models that possess high interpretability and are capable of extracting nonlinear features from time series data are particularly appealing to financial market participants, in particular those in quantitative finance. Signature is a good synthesis of ordered stream data (High-frequency intraday financial data is often regarded as continuous streams) that can capture their nonlinear features with clear expressions. The concept of path signature is initially introduced and developed by Lyons \citep{Lyons1998}, see also Lyons and Qian \citep{lyons2002system} in the context of rough path analysis, which has become one of the core tools in the analysis of stream data. The key idea in rough path theory is that information contained in complex dynamic systems can be characterized fully by signature. The signature of a continuous path $X$ is the sequence of its iterated tensor integrals, so that it maps a continuous path $X$ to a unique and complete feature representation $S(X)=(1,X^1,X^2,\cdots)$, where $X^n$ is the $n$-th order iterated integral (for $n=1,2,\cdots$). This map retains important geometric and topological information, thereby gives a complete and efficient description of the path $X$. The signature of the path $X$ gives us all information needed for determining nonlinear functions of the path $X$ via Taylor's expansion, thereby providing a means of extracting higher-order nonlinear features, obtaining information needed for propose of modeling. Therefore, signatures are capable of effectively capturing both locally and globally important features of data streams, including correlations and path dependencies, reflecting similarities and differences between components of stream data.

These properties of signatures yield their great potential in extracting high-dimensional features in complex and high-frequency data. Gyurko et al. \citep{Gyurko2014} used the signature method to classify financial data flows and successfully distinguish characteristics of market behavior in different time periods. They found that the signature method can capture subtle changes in market data, such as volume distribution and price volatility patterns. Kalsi et al. \citep{Kalsi2020} proposed a signature method to solve the optimal execution problem. These authors have used a truncated signature to give meaningful approximations of the optimal transaction speed. Signature method has been used effectively to solve path-dependent problems and has been proved to be superior in feature extraction, such as for American option pricing \citep{Bayer2023}, time series generation \citep{Ni2021}, and etc. These successful applications of the signature method are all based on three key elements of the mathematical construction of the signature. First, the signature index has time translation invariance, which naturally adapts to the problems of sampling frequency difference and baseline drift of financial time series. Second, the higher order signature can be used to the study of nonlinear features of stream data, such as curvature and wave mode, which breaks the limitation of the traditional spread model relying only on the first moment. Third, a discrete price stream $\{(t_i,X_{t_i})\}$ can be transformed into a continuous path by the Lead-Lag transformation, so that signature features can be extracted.

Nevertheless, the application of signature in trading strategies has not been systematically studied. Although some effort has been made so far to combine signature with LSTM, Transformer and other models for time series analysis \citep{Levin2013,Chevyrev2016}, these studies mostly focus, however, on single-path prediction, but not on joint paths. In addition, the traditional truncated signature is limited due to the well-known curse of dimensionality: the number of $n$-order signature features for $d$-dimensional path increases by $O(d^n)$ level, leading to high computational complexity when the signature method is applied to multi-asset portfolio directly. Moreover, while the mathematical formulation of the signature is elegant, its interpretation within financial markets remains somewhat abstract. This lack of concrete intuition can hinder its practical utility as an indicator for trading decision. In our view, how to reasonably construct feature extraction indexes based on signature method and scientifically apply them in statistical arbitrage models is the key to optimize the trading strategy.

Motivated by the issues above, we innovate the signature method and propose a novel path feature named as segmented signature in this paper. By using a decomposition of the original signature, we extract segmented signature (a kind of transformation of L{\'e}vy area)  for reflecting the interactive information and trend dynamics of multiple financial asset price sequences. The segmented signature, which possesses good interpretability and stability, can be seen as an effective nonlinear feature of path interactivity. According to this property, segmented signatures can be used naturally in pair trading, which leads to a strategy of pair trading capturing the complex relationships of assets. We use segmented signature as a filter in the pair trading strategy, which enhances the precision of the decision-making process, thus that validate the feasibility of segmented signature. In our empirical study, it is found that, besides segmented signature separated from the original signature, the decomposed term product of path difference is also meaningful and indeed helpful, representing whether paths change in the same direction or not. With a careful analysis together with numerical experiments, we may propose a double filters strategy SE-SIG-DIFF based on segmented signature and their covariation of increments (product of one-order difference). A relative comprehensive empirical research based on this idea is carried out in our paper. The empirical results show that the new SE-SIG-DIFF strategy has significant effectiveness in improving Sharpe ratio, increasing returns, controlling maximum drawdown and other aspects.

Our study contributes to existing research in several aspects. First, we have proposed the segmented signature which is an effective and interpretable indicator describing the path interactivity. It is worth noting that the segmented signature has low computational complexity and inherent dimension, so it can be easily calculated and used as a good indicator for both individual and institutional investors who are interested in quantitative trading. Second, as a feature or as an indicator, segmented signature is calculated from its own path and does not require additional information, making it an ideal indicator for trading. Moreover, the original signature seems to perform poorly in trading strategy, and it is proposed in our paper to decompose it into two separate indicators: segmented signature and covariation of increments component. This kind of decomposition, though mathematically trivial, in fact extracts mixed information into more independent and characteristic information, and therefore is very helpful for quantitative strategies. The decomposed indicators used as double filters will greatly improve the effectiveness of pair trading strategies. We believe that the method we have proposed is also interesting to researchers in mathematical modeling and rough path theory, as well as scholars and investors interested in trading arbitrages.

The remaining part of the paper is organized as follows. In Section 2, we introduce the theory of signature and decompose it to obtain the segmented signature. Then, combined pair trading with segmented signature, we propose the SE-SIG-DIFF strategy, as presented in Section 3, along with the benchmark pair trading strategy and required data sources. Comprehensive empirical study and statistical test are conducted in Section 4, validating the effectiveness and feasibility of our new trading strategy. In Section 5, we present the conclusions and outlook of our work.

\section{Decomposition of signature}
\subsection{Theory of signature method}
In this subsection, we present a concise overview of the signature, including its core concepts and essential properties. For a comprehensive treatment of the theoretical foundations, we refer interested readers to the papers \citep{lyons2002system} and \citep{lyons1991}. Now we introduce some notations. Let $\mathbb{R}^d$ be the $d$-dimensional Euclidean space, $\mathcal{V}^p(J,E)$ be the set of continuous path $X:J \rightarrow E$ of finite p-variation, where $J$ is the parameter space and $E$ is the range space.

Let $J$ be a compact interval and $X\in \mathcal{V}^p(J,\mathbb{R}^d)$ such that the following integration makes sense. The signature $S(X)$ of $X$ over the time interval $J$ is defined as an infinite series of $X_J^n$, i.e. $S(X)_J=\left(1,X_J^1, \cdots,X_J^n, \cdots \right) $, where

$$X_J^n=\idotsint \limits_{t_1 <\cdots <t_n;\ t_1\cdots t_n \in J} \mathbf{d}X_{t_1} \otimes \cdots \otimes \mathbf{d}X_{t_n} \in (\mathbb{R}^{d})^{\otimes n},$$
where $t_1\cdots t_n \in J$ is the time division. The notation $\otimes$ means that the integration is defined in the sense of tensor product. Let $S_n(X)_J$ denote the truncated signature of $X$ of degree $n$, i.e. $S_n(X)_J=(1,X^1_J,\cdots,X ^n_J )$.  Actually, the signature is an important geometric feature of the original path. The first order terms of the signature are the increments of the paths and the second order terms are related to the area enclosed by two paths. Low-order signature can be seen as a projection of high-order signature. 

Let $X \in \mathcal{V}^1 (J,E)$. Then $S(X)$ determines $X$ up to the tree-like equivalence \citep{Hambly2010}. A tree-like path refers to a path that retraces itself in such a way that its trajectory is entirely canceled out. The precise definition of tree-like equivalence is provided in \citep{Hambly2010}. Although we do not delve into the formal details of this equivalence, the corresponding relationship ensures that the signature of a path is, in a certain sense, unique.

In practical applications, working with the full signature is computationally infeasible. Due to finite-precision constraints in digital computing, we must instead employ the truncated signature as previously defined. The full signature offers a complete characterization of a path, but truncation inevitably discards higher-order terms, potentially leading to information loss. However, Terry \citep{lyons2006} mentioned the information attenuation property of signature. Specifically, assume that $X$ is the $d$-dimensional path with bounded variation, we can obtain
$$\left \Vert \;\; \idotsint \limits_{t_1 <\cdots <t_n;\ t_1\cdots t_n \in J} \mathbf{d}X_{t_1}^{(i_1)} \cdots \mathbf{d}X_{t_n}^{(i_n)} \right \Vert \leq \frac{\Vert X \Vert_1^n}{n!}, \quad 1 \leq i_1,\cdots i_n \leq d,$$
with
$$\Vert X \Vert_1^n=\sup \limits_{\{t_i\} \in J}\sum_{i} \vert X_{t_{i+1}}-X_{t_i} \vert.$$
This property establishes that the higher-order terms of a signature decay at a factorial rate. Consequently, truncating the signature by retaining only its initial terms leads to minimal information loss, as the discarded higher-order terms contribute negligibly. This enables the truncated signature to serve as a highly effective representation of the path, making it a powerful feature in the analysis of path-dependent data. This is the primary reason we regard the signature as a fundamental path-based descriptor for capturing salient features in high-frequency data.

In practice, for the medium and high-frequency financial data which is chaotic, dynamic and complex, signature technique exhibits the ability to extract structural features. By mapping raw data sequences into feature space, it facilitates the extraction of more detailed information and features from the original data. These crucial pieces of information, although important, still retain a certain degree of abstraction and confusion in terms of their explainability and practical significance within the context of real life trading and financial markets. We reiterate the notation used in the following contents: $X_t^{(i)} \in \mathbb{R}^d$ is the value $i$-th dimension of path $X$ at time $t$, $X_{s,t}^n \in (\mathbb{R}^{d})^{\otimes n}$ means $n$-order signature in time interval $[s,t]$. For example, when $d=2$, the second order signature $X_{s,t}^2$ is a $2\times2$ matrix with the component $X_{s,t}^{(i,j)}$, $i,j \in 1,2$. In the notation used in this article, a single time subscript ($X_t^{(i)}$) indicates the value of a path at a specific time, while two time subscripts ($X_{s,t}^{(n)}$) signify the $n$-order signature.

\subsection{Segmented signature}

Our motivation for proposing the segmented signature stems from the limitations observed in the direct application of the original signature as a feature or indicator. According to the definition and labels in subsection 2.1, the first-order signature $X_t-X_s \in \mathbb{R}^d$ is actually the path difference, corresponding to the price increment between time $s$ to $t$ in financial market. The most commonly used second-order signature matrix $X_{s,t}^2 \in \mathbb{R}^{d \times d}$ contains a lot of information that is vague and abstract. From a mathematical perspective, some kind of transformation of it represents the L\'evy area of paths (mentioned later). However, in finance, it can only be interpreted as an approximate measure of the path's complexity. In other words, original signature is not suitable to be considered as a feature or signal directly because the strategies based on it are ambiguous. So, we try to decompose the signature and explore an explicable and clear path feature to facilitate the financial decision.

\begin{figure}[H]
    \centering
    \includegraphics[width=0.4\textwidth]{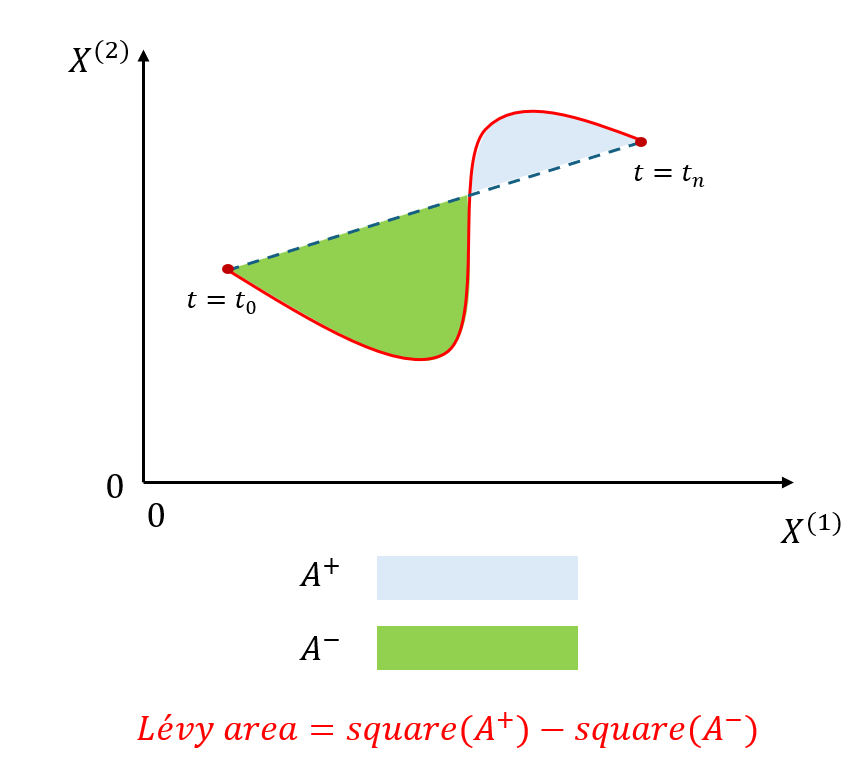}
    \caption{L\'evy area between $X^{(1)}$ and $X^{(2)}$ (Simple case)}
    \label{Levy1}
\end{figure}

\begin{figure}[H]
    \centering
    \includegraphics[width=0.6\textwidth]{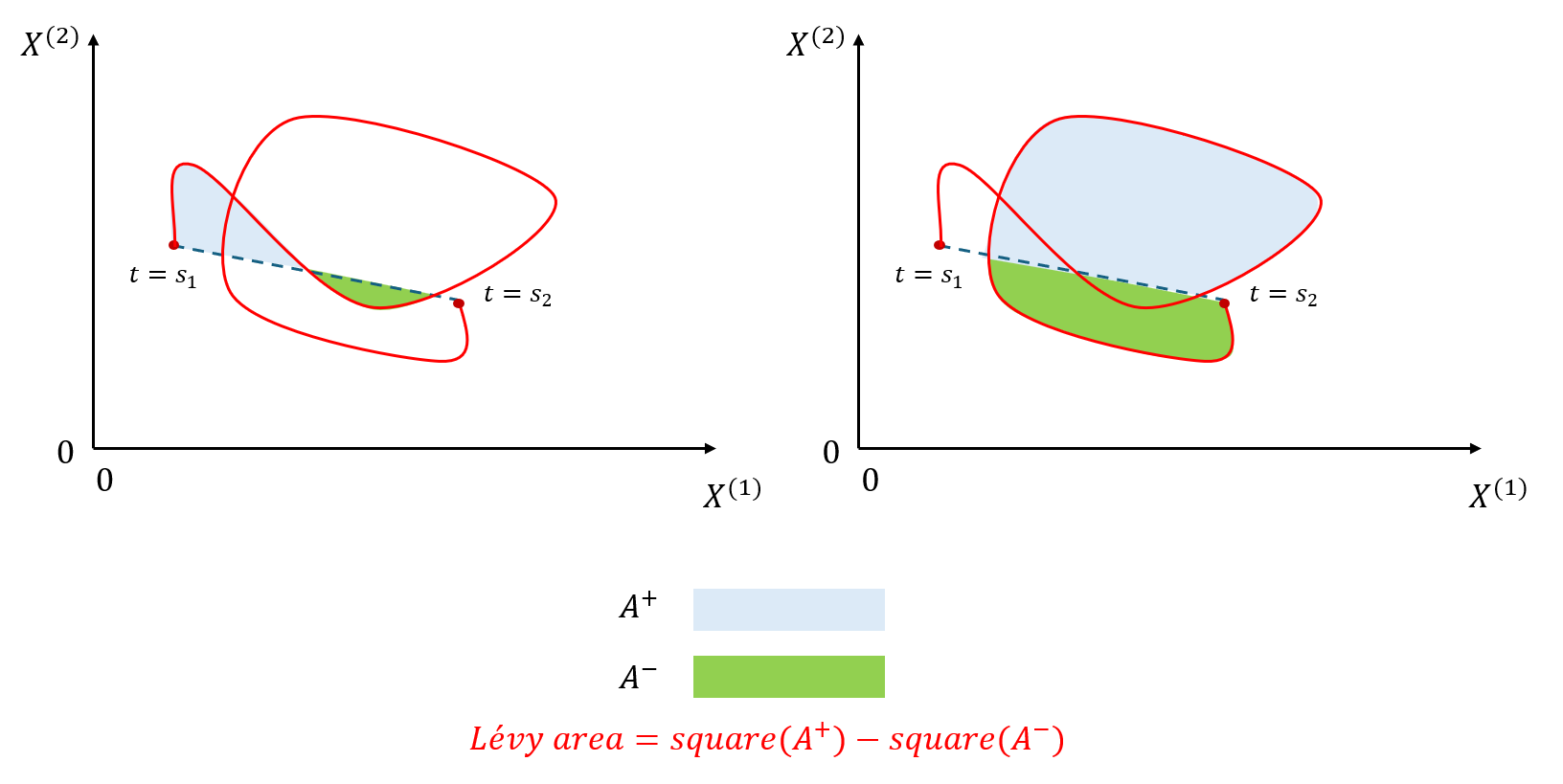}
    \caption{L\'evy area between $X^{(1)}$ and $X^{(2)}$ (Complex case)}
    \label{Levy2}
\end{figure}

In this subsection, we introduce the decomposition of the original signature and explain why the decomposed and processed signature has better interpretability and more intuitive reflection of path characteristics. Because of the information attenuation property, it is convincing that second-order signature can contain enough path information, thereby high-order signature is not necessary in practice. Taking 2-dimensional path as an example, we first explain the significance of the second-order signature matrix $X_{s,t}^{2}$ with the component $X_{s,t}^{(i,j)}$, , which can be calculated by the following expression:

$$X_{s,t}^{(i,j)}=\int_{s<u_1<u_2<t} \text{d}X_{u_1}^{(i)} \text{d}X_{u_2}^{(j)},$$
where $i,j=1,2$, $[s,t]$ is time interval. According to the relationship:
$$X_{s,t}^{(i,j)}+X_{s,t}^{(j,i)}=X_{s,t}^{(i)}X_{s,t}^{(j)},\quad X_{s,t}^{(i,i)}=\frac{(X_{s,t}^{(i)})^2}{2}.$$
Then $X_{s,t}^{2}$ can be decomposed to $A_{s,t}^2$ plus $D_{s,t}^2$, expressed in terms of matrix components:
$$X_{s,t}^{(i,j)}=A_{s,t}^{(i,j)}+D_{s,t}^{(i,j)},$$
where
$$A_{s,t}^{(i,j)}=\frac{1}{2}(X_{s,t}^{(i,j)}-X_{s,t}^{(j,i)}), \quad D_{s,t}^{(i,j)}=\frac{1}{2}X_{s,t}^{(i)}X_{s,t}^{(j)}.$$ 
$A_{s,t}^{(i,j)}$ and $D_{s,t}^{(i,j)}$ are the component of matrix $A_{s,t}^2$ and $D_{s,t}^2$ respectively \citep{Levin2013}. $A_{s,t}^{(1,2)}$ term is L\'evy area, representing somewhat relationships of paths which is what researchers mainly concern, and $D^2$ reflects the product of paths increments. It is precisely because it contains a lot of complex path information that we believe it will have good effects on extracting the path features. Then, we try to think about how to understand and explain $A_{s,t}^{(1,2)}$. The L\'evy area $A_{s,t}^{(1,2)}$ is the enclosed area of a trajectory of two-dimensional path and its chord (Figure \ref{Levy1} and \ref{Levy2}).

\begin{figure}[H]
    \centering
    \includegraphics[width=0.4\textwidth]{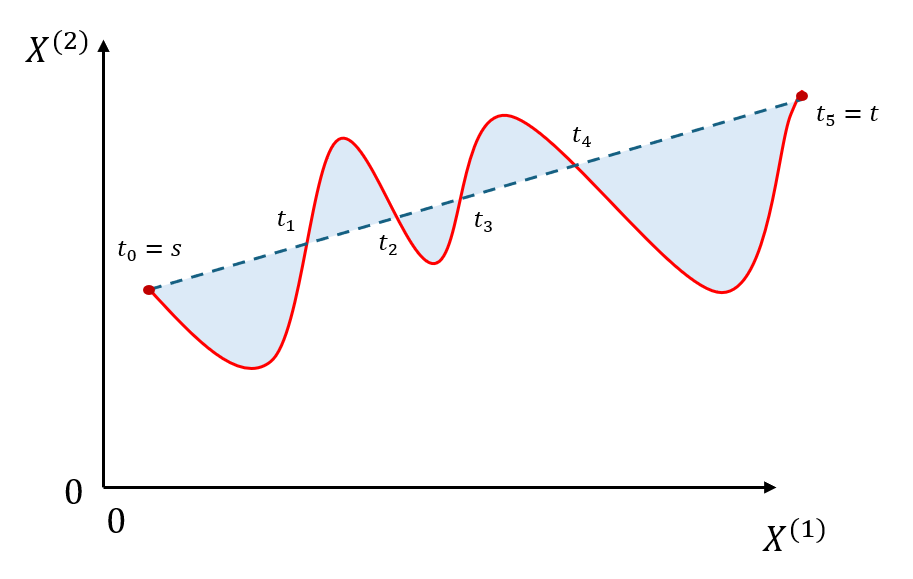}  
    \caption{Segmented L\'evy area between $X^{(1)}$ and $X^{(2)}$}
    \label{Segmented}
\end{figure}

L\'evy area is the sum of $A^{+}$ minus $A^{-}$. From an intuitive perspective, the L\'evy area appears to reflect the degree of interaction or correlation between paths. For instance, when two assets exhibit stronger correlation, their interactions seem more aligned, potentially resulting in a smaller L\'evy area. If two assets are absolutely linearly correlated, then the line connecting their paths is just the chord between the initial and terminal points. At this time, L\'evy area $A_{s,t}^{(1,2)}$ is 0, thus it might be abstractly regarded as representing a certain type of nonlinear relationship of two paths. However, the opposite is not true. When L\'evy area $A_{s,t}^{(1,2)}$ is 0, one possible scenario is that $A^{+}$ and $A^{-}$ (ref Figure \ref{Levy1} or \ref{Levy2}) completely cancel each other out, making it impossible to determine the degree of association between the two assets. To address this issue, we introduce a novel metric, termed the segmented signature matrix $C_{s,t}^{2}$, which is the absolute value of the L\'evy area over each segment. The component of segmented signature feature $C_{s,t}^{2}$ is $C_{s,t}^{i,j}$, which is defined as follows:

\begin{align*}
    C^{(i,j)}_{s,t}=\sum_{r=0}^{n-1} |A^{(i,j)}_{t_r,t_{r+1}}|,
\end{align*}
where $t_r,r=1,\cdots n$ is a time division, $t_0=s,t_n=t$. The $t_r$ are selected as the crossing of chord and the path, shown in Figure \ref{Segmented}. In this way, we divide the whole time interval into different interval and calculate the absolute value of L\'evy area (shaded area in Figure \ref{Segmented}).

\begin{figure}[H]
    \centering
    \includegraphics[width=1\textwidth]{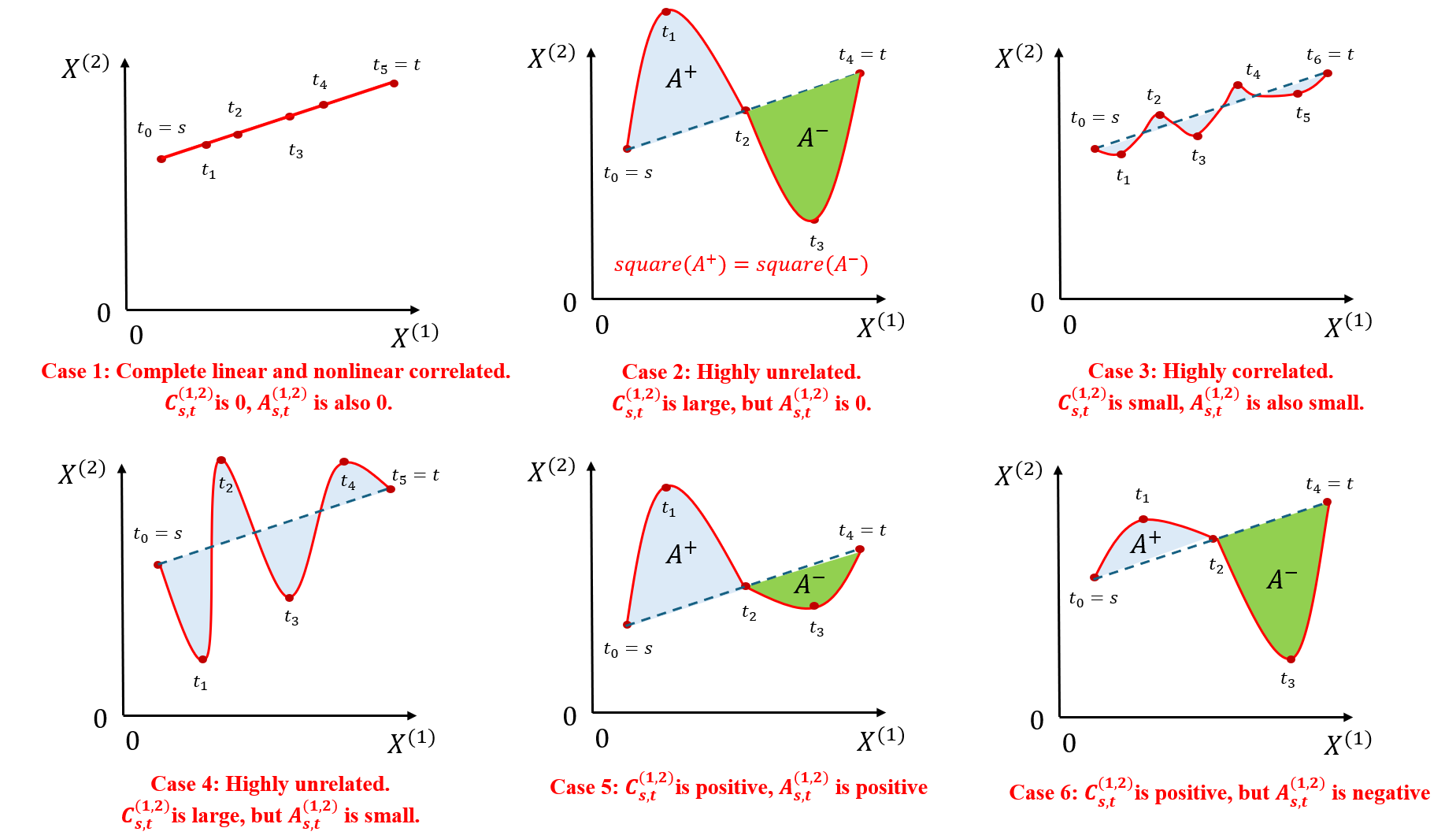}  
    \caption{Some Cases Described by $A_{s,t}^{(1,2)}$ and $C_{s,t}^{(1,2)}$}
    \label{levycase}
\end{figure}

There are several advantages that motivate us to explore this new feature of paths and we plot the Figure \ref{levycase} to give an intuitive explanation of this matter. Let us assume that the $X^{(1)}$ and $X^{(2)}$ are the two asset paths in financial market. First, when second-order segmented signature $C_{s,t}^{(1,2)}$ is zero, we can say exactly that the two assets are completely linearly correlated. However, $A_{s,t}^{(1,2)}$ does not necessarily imply this (Figure \ref{levycase}: Case 1 and Case 2). Second, it is convincing that segmented signature reflects the correlation, or interactivity of paths, with smaller values indicating greater interactivity. However, the original second-order signature $X_{s,t}^{(1,2)}$ (or L\'evy area $A_{s,t}^{(1,2)}$) does not have the above properties (Figure \ref{levycase}: Case 3 and Case 4). Additionally, segmented signature is always positive, but the sign of $A_{s,t}^{(1,2)}$ is difficult to explain (Figure \ref{levycase}: Case 5 and Case 6). Therefore, we believe that compared with original signature, segmented signature filters out part of the information unrelated to the path interaction, more directly reflects the path interaction, and has better interpretability, indicating that it is more suitable as a nonlinear trading signal.

Reviewing this subsection, we introduce the motivation and the processing of decomposition of second-order signature. It is particularly important to note that after decomposing $X_{s,t}^2$, getting $A_{s,t}^2$ and optimizing it to $C_{s,t}^2$, we also obtain a matrix $D_{s,t}^2$. This matrix $D_{s,t}^2$, a component often overlooked by researchers, proves to be highly significant based on our subsequent empirical findings in section 4. As the sign of $D_{s,t}^{(1,2)}$ indicates whether the two variables move in the same direction, signifying a positive or negative correlation in financial market. We call $D_{s,t}^{(1,2)}$ as covariation of increments. This section employs a two-dimensional example to elucidate the theory of signatures, which corresponds to our subsequent pairs trading strategy involving two assets. The theoretical framework remains entirely consistent in higher-dimensional cases. Indeed, in the two-dimensional setting, for matrices $A_{s,t}^2$, $C_{s,t}^2$, and $D_{s,t}^2$, we primarily utilize their components $A_{s,t}^{(1,2)}$, $C_{s,t}^{(1,2)}$, and $D_{s,t}^{(1,2)}$.


\section{Methodology}

\subsection{Pair trading, data and parameter setting}
According to the signature theory in section 2, segmented signature could reflect the interactivity between two assets. As is well known, one of the traditional trading strategies referring to the relationship between two assets is pair trading. Pair trading is a market-neutral trading strategy that exploits the relationship between the price movements of two related assets. When the price difference between two assets deviates from its historical average, traders can go long on one asset and short on the other asset at the same time, expecting the price difference to return to its mean level. This strategy is based on the principles of statistical arbitrage and achieves returns by looking for pairs of co-integrated assets. Based on the compatibility of signature and pair trading, pair trading is regarded as the benchmark to verify the effectiveness of signature on strategy in the real futures market.

Our data includes futures minute-level data from November 1, 2024 to December 31, 2024 in the Chinese market and the United States market \footnote{The data comes from RQData and CME dataset, and its API is provided by OXFORD SUZHOU CENTRE FOR ADVANCED RESEARCH.}. Assets are paired for pair trading. Assume that $X^{(1)}_t$ and $X^{(2)}_t$ represent the price of two assets, and we take log to stabilize the time series of price data. In pair trading theory, statistics $Z_{t}$ is defined to identify trading signals, the $Z_{t}$ is calculated in each rolling window $w$ as:

$$Z_{t} = \frac{(S_t - \mu_t)}{\sigma_t}, $$
where $S_t$ represents the log price spread between assets $X^{(1)}_t$ and $X^{(2)}_t$ after balancing ($S_t = \log(X^{(1)}_t) - \beta \log(X^{(2)}_t)$), where $\beta$ is the number of lots which is the regression coefficient of $X^{(1)}$ and $X^{(2)}$, ensuring risk-neutral positioning.
$\mu_t$ and $\sigma_t$ represent the rolling mean and rolling standard deviation of spread at time window $t-w$ to $t$. In theory, there is a threshold $Z_{\text{score}}$ that the normalized statistics $Z_{t}$ should be in the interval $[-Z_{\text{score}},Z_{\text{score}}]$, if $Z_{t}$ exceed this interval, we note that the a low-probability event occurs, indicating a deviation in the price spread. In our subsequent empirical experiments in subsection 4.2, we set $w=60, Z_{\text{score}=2}$ and other choices of parameters combination are included in statistical test. Trading signals are generated when the $Z_{t}$ exceeds $\pm Z_{\text{score}}$: a long position is taken on the under-performing asset while shorting the outperforming one. Positions are closed when the $Z_{t}$ reverts to zero. This implementation follows the statistically rigorous framework in \citep{vidyamurthy2004pairs,stempien2025hybrid,chen2019empirical} and the code framework of pair trading strategy was shown in Algorithm \ref{al1}.

In our framework, we account for transaction costs by deducting a round-trip cost of $0.05\%$  of the notional value for each completed trade. This is a common conservative estimate of costs in quantitative research in the Chinese futures market. Specifically, round-trip cost includes exchange fees (0.00015\% to 0.008\% ), brokerage markups (0.001\% to 0.012\%) and slippage (0.01\%–0.03\%).  For the sake of conservatism, researchers often use 0.05\% as the benchmark for deducting costs from each return, so 0.05\% is a biased, robust, and widely accepted empirical value in Chinese futures market.

And for American market, a realistic and widely used estimate for total transaction costs in U.S. commodity futures research is 0.01\% to 0.02\% of notional value per round-trip trade. However, for commodity futures with liquidity differentials (such as some agricultural futures), the conservative round trip comprehensive cost (in nominal value) may reach 0.02\% to 0.04\%. Therefore, for the US market, we also use a more conservative 0.05\% as the total transaction cost rate.

As for risk management, to control downside risk, in addition to the signal generation logic outlined in Algorithm\ref{al1}, we impose a stop-loss rule that limits the maximum loss per trade to the maximum drawdown of the strategy performance observed during the two-month period before strategy period. This threshold is fixed and applied uniformly throughout the strategy period. We note that while dynamic volatility scaling is widely used in portfolio optimization \citep{slusarczyk2025optimal}, our approach employs a fixed stop-loss rule based on pre-strategy maximum drawdown to ensure simplicity, transparency, and robustness in high-frequency execution environments.

\begin{algorithm}[H]\label{al1}
\SetAlgoLined
\caption{Pair Trading Strategy}
\SetKwInOut{Input}{Input}
\SetKwInOut{Output}{Output}
\SetKwIF{If}{ElseIf}{Else}{if}{:}{elif}{else}{end}

\Input{
    $X^{(1)} = (X^{(1)}_1, \dots, X^{(1)}_T)$: price series of asset 1 \\
    $X^{(2)} = (X^{(2)}_1, \dots, X^{(2)}_T))$: price series of asset 2 \\
    $w$: rolling window size (e.g., 60 minutes) \\
    $Z_{\text{score}}$: threshold for $Z_t$ (e.g., 2)
}

Let $\mathcal{L}^{(1)} = (\log(X^{(1)}_1), \dots, \log(X^{(1)}_T))$\;
Let $\mathcal{L}^{(2)} = (\log(X^{(2)}_1), \dots, \log(X^{(2)}_T))$\;

Fit regression: $\mathcal{L}^{(1)} = \alpha + \beta \mathcal{L}^{(2)} + \epsilon$\;
Obtain optimal hedge ratio $\beta$\;

\For{$t = 1$ \KwTo $T$}{
    $S_t = \mathcal{L}^{(1)}_t - \beta \cdot \mathcal{L}^{(2)}_t$\;
}

\For{$t = w+1$ \KwTo $T$}{
    Compute rolling mean: $\mu_t = \frac{1}{w} \sum_{i=t-w}^{t-1} S_i$\;
    Compute rolling std: $\sigma_t \gets \sqrt{\frac{1}{w-1} \sum_{i=t-w}^{t-1} (S_i - \mu_t)^2}$\;
    Compute $Z_t$: $Z_t \gets \frac{S_t - \mu_t}{\sigma_t}$\;

    \If{$Z_t > Z_{\text{score}}$}{
        \KwRet signal = (short $X^{(1)}$, long $X^{(2)}$)\;
    }
    \ElseIf{$Z_t < -Z_{\text{score}}$}{
        \KwRet signal = (long $X^{(1)}$, short $X^{(2)}$)\;
    }
    \Else{
        \KwRet signal = (hold, hold)\;
    }
}
\end{algorithm}

\subsection{Selection of assets}
In this section, we illustrate the idea and technology of our assets selection. In generally, we adopt a combination of subjective judgment and objective analysis for asset selection. Specifically, we first define the asset universe based on experience and historical data. Since pair trading requires a certain degree of correlation between assets, the selected assets must belong to the same broad category. For example, we assume that precious metals exhibit some degree of correlation, therefore, we select assets from precious metal futures to construct a asset group. When choosing specific assets, statistical methods are employed to assist us in the selection of specific futures varieties. We consider an asset pair suitable for pair trading only if it exhibits a sufficiently high correlation and if the price spread demonstrates a strong mean reversion characteristic. Taking metal futures as an example, we provide a detailed explanation of our asset selection process in the following.

For Chinese metal futures market, we obtain the metal futures data: AU(gold futures), AG(silver futures), SN(tin futures), AL(aluminum futures), CU (copper futures), NI (zinc futures), which are all main and active futures products. First, futures with a correlation coefficient greater than 0.5 can be regarded as having a relatively strong correlation. We fix the window size and calculate the Pearson correlation coefficient matrix for each pair of assets, shown in Table \ref{correlation}.

\begin{table}[H]
\centering
\caption{Pearson Correlation Coefficient Matrix for Metal Futures}
\[
\begin{array}{c|cccccc}
\hline
   & \text{AU} & \text{AG} & \text{CU} & \text{AL} & \text{SN} & \text{NI} \\
\hline
\text{AU} & 1.0000 & 0.7693 & 0.1990 & 0.5895 & 0.5654 & 0.3539 \\
\text{AG} & 0.7693 & 1.0000 & 0.5546 & 0.6704 & 0.6581 & 0.6869 \\
\text{CU} & 0.1990 & 0.5546 & 1.0000 & 0.7393 & 0.6754 & 0.4879 \\
\text{AL} & 0.5895 & 0.6704 & 0.7393 & 1.0000 & 0.9410 & 0.7567 \\
\text{SN} & 0.5654 & 0.6581 & 0.6754 & 0.9410 & 1.0000 & 0.7128 \\
\text{NI} & 0.3539 & 0.6869 & 0.4879 & 0.7567 & 0.7128 & 1.0000 \\
\hline
\end{array}
\]\label{correlation}
\end{table}
According to our standard, the correlation coefficient between NI and the other two types of futures is less than 0.5. Therefore, we first exclude NI from consideration. The other key condition for our pair trading strategy to make profit is that the price spread should have a strong mean reversion characteristic. If the price spread does not exhibit mean-reverting behavior, deviations from the equilibrium may persist indefinitely, making it impossible to capture profitable trading opportunities. To achieve this, we introduce the Hurst exponent in pair trading strategies, which is firstly proposed by Hurst \citep{Hurst1951}.

Hurst exponent is a statistical measure that is used to study scaling properties in time series. Generalized Hurst Exponent(GHE) verifies whether some statistical properties of data adjust with both number of observations and the time resolution, the readers who are interested in the use of GHE in trading strategy can refer to \citep{bui2022applying}. If we have the time series $X(t)$, $t=1,2,\cdots n$, we can define the statistic $E_p(s)$ according to the following formula:

$$E_p(s)=\frac{\mathrm{mean}(|X(t+s)-X(t)|^p)}{\mathrm{mean}(|X(t)|^p)},$$
where $s$ varies between 1 to $s_{max}$, while $s_{max}$ is chosen as a quarter of the length of the series. The GHE characterizes the scaling properties of $X(t)$ and therefore is associated with the scaling behavior of
the statistic $E_p(s)$. According to the power-law, the GHE $H(q)$ is calculated as:

$$E_p(s) \propto s^{qH(q)} \rightarrow \log(E_p(s)) \propto q\log(s)H(q),$$
where $\propto$ represents direct proportionality. So taking $\log(E_p(s))$ as dependent variable y and $q\log(s)$ as independent variable x, the GHE $H(q)$ is the regression coefficient of the linear regression between y and x. We select $q=1$, which is close to the original Hurst exponent, and people who are interested in investigating long-range dependence can set $q=2$. The Hurst exponent between 0.5 and 1 indicates momentum behavior and between 0 and 0.5 indicates mean-reverting behavior. Reader who are interested in the detailed method can refer to \citep{bui2022applying}. So in our pair trading task, we hope that the Hurst exponent of our price spread between futures is less than 0.5, and the lower, the better. So we calculate the Hurst exponent for the price spread (after balancing) of each asset pair in a window size, the results of mental futures are shown in Table \ref{hurst}.

\begin{table}[H]
\centering
\caption{Hurst Exponent of Price Spread in Metal Futures Group}
\resizebox{\textwidth}{!}{
\begin{tabular}{c|cccccccccc}
\hline
   & AU\&AG & AU\&CU & AU\&AL & AU\&SN & AG\&CU & AG\&AL & AG \&SN & CU \& AL & CU \& SN & AL \& SN\\
\hline
Max \ value & 0.4642 &0.6785 & 0.4630 & 0.4634 & 0.6740& 0.4679 &0.4593 &0.5459&0.5559&0.4578\\
Min\ value & 0.0142 & 0.1224 & 0.0725 & 0.0804 & 0.1166 & 0.0952  &0.0169 &0.1118&0.1183&0.0794\\
Mean\ value & 0.2168 & 0.3919& 0.2111 & 0.2220 & 0.2139 & 0.2049  &0.2138 &0.2976&0.2976&0.2052\\
\hline
\end{tabular}
}
\label{hurst}
\end{table}

\begin{figure}[H]
    \centering
    \includegraphics[width=1\textwidth]{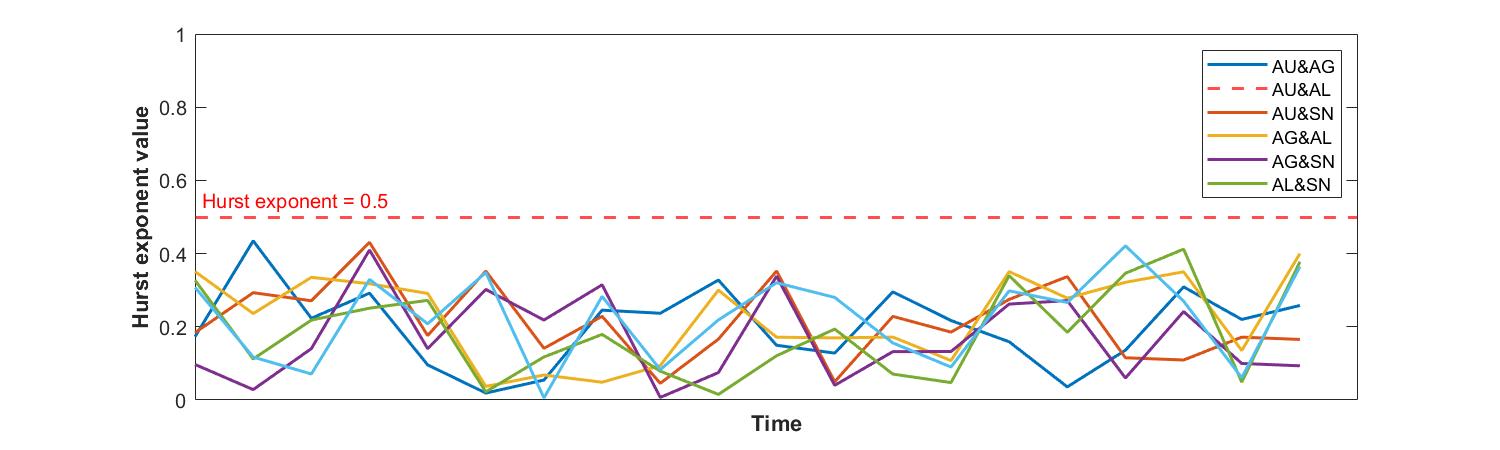}
    \caption{Hurst Exponent Time Series of Selected Metal Futures}
    \label{hurstpic}
\end{figure}

From Table \ref{hurst}, we find that some Hurst exponents of the price spread between other futures and Copper futures are greater than 0.5, which means that Copper futures are not suitable for pair trading combined with other assets. So after this process, we select the AU, AG, AL and SN futures from the metal futures groups as their price spread Hurst exponent is always below 0.5. We plot the time series of the Hurst exponent for the future pair in the metal group, shown in Figure \ref{hurstpic}.

Obeying to the steps behind, five groups of targeted futures are selected and shown as follows. In Chinese futures market, we select three groups of futures:

\begin{tikzpicture}
  \draw[fill=black] (0,0) circle (0.1cm); 
\end{tikzpicture} Gruop 1 (Metal futures): AU(gold futures), AG(silver futures), SN(tin futures), AL(aluminum futures).

\begin{tikzpicture}
  \draw[fill=black] (0,0) circle (0.1cm); 
\end{tikzpicture} Group 2 (Agricultural product futures): C(corn futures), B(soybeans futures), CF(cutton futures), M(soybean meal futures).

\begin{tikzpicture}
  \draw[fill=black] (0,0) circle (0.1cm); 
\end{tikzpicture} Group 3 (Oil related product futures): MA(methanol futures), SC(crude oil futures), Y(soybean Oil futures), RB(rebar futures).

To demonstrate the universality and broad applicability of our method, we have included an analysis of U.S. futures market data. We have prefixed the full names of US futures varieties with "US" to avoid redundant symbols. Referring to the futures varieties of the Chicago Mercantile Exchange, after the same selecting steps, we choose the following two groups of futures:

\begin{tikzpicture}
  \draw[fill=black] (0,0) circle (0.1cm); 
\end{tikzpicture} Group 4 (US Metal futures): GC(US gold futures), SI(US silver futures), PA(US palladium futures).

\begin{tikzpicture}
  \draw[fill=black] (0,0) circle (0.1cm); 
\end{tikzpicture} Group 5 (US Agricultural product futures): ZC(US corn futures
), ZW(US wheat futures), ZS(US soybeans futures), ZL(US soybean oil futures).

\subsection{Signature trading strategy}

In this subsection, we introduce our trading strategy based on segmented signature and other benchmark strategies for comparison with our strategy. As mentioned in subsection 2.2, the smaller the values of the signature components $C_{s,t}^{(1,2)}$ of the two assets, the stronger their nonlinear correlation or interaction. Therefore, the motivation of our new strategy is that in pair trading, a relatively robust arbitrage opportunity is identified only when both linear and nonlinear correlation signals between the two assets are present, accompanied by a price deviation. So in our strategies, a trade is considered for triggering only when current signature value falls below a (certain) threshold, we set it as the historical mean of the signature, which avoids look-ahead bias and brings more robustness \citep{stempien2025hybrid}. In addition, since the selected assets belong to the same category, the directional signal $D_{s,t}^{(1,2)}$ is expected to be positive. That is, when the two price paths are positively correlated and move in the same direction, the price deviation is likely to represent a genuine arbitrage opportunity.

In summary, the signature and segmented signature are used as filters to select trading opportunities that do not meet our specified criteria. Below are four different signature strategies that we used in the empirical experiments, each of them are separately used to compare its effectiveness in the real futures data.

\begin{tikzpicture}
  \draw[fill=black] (0,0) circle (0.1cm); 
\end{tikzpicture} Normal Pair Trading (No SIG): Traditional pair trading method without any filters or signals.

\begin{tikzpicture}
  \draw[fill=black] (0,0) circle (0.1cm); 
\end{tikzpicture} Pair Trading with original signature (SIG): Traditional pair trading method with original second-order signature component $A_{s,t}^{(1,2)}$ as filter.

\begin{tikzpicture}
  \draw[fill=black] (0,0) circle (0.1cm); 
\end{tikzpicture} Pair Trading with segmented signature (SE-SIG): Traditional pair trading method with second-order segmented signature component $C_{s,t}^{(1,2)}$ as filter.

\begin{tikzpicture}
  \draw[fill=black] (0,0) circle (0.1cm); 
\end{tikzpicture} Pair Trading with segmented signature and path difference product (SE-SIG-DIFF): Traditional pair trading method with second-order segmented signature component $C_{s,t}^{(1,2)}$ and covariation of increments $D_{s,t}^{(1,2)}$ as double filters.

Specifically, we use the segmented signature and the product of differences as filters to guide the investment. The SIG strategy is adding second-order signature threshold as a condition to filter transactions. SE-SIG strategy changes condition from second-order signature threshold to second-order segmented signature threshold. Based on SE-SIG strategy, SE-SIG-DIFF strategy adds path difference product as another condition. Clearly, SE-SIG-DIFF strategy is the most complex strategy compared to other three strategies, so we present the code framework of SE-SIG-DIFF strategy in Algorithm \ref{se-sig}.

\begin{algorithm}[H]\label{se-sig}
    \SetAlgoLined
    \caption{Segmented signature + path difference strategy}
    \SetKwInOut{Input}{Input}
    \SetKwInOut{Output}{Output}
    \SetKwIF{If}{ElseIf}{Else}{if}{:}{elif}{else}{}
    
    \Input{
        $X^{(1)} = n \times 1$, value vector of the first futures \\
        $X^{(2)} = n \times 1$, value vector of the second futures \\
        $\alpha$ = Summary of parameters (including window size, initial asset and so on) \\
        $i=1,2,\ldots,T$ is the date.
    }
    
    \SetKwFunction{Signature}{Signature}
    \SetKwFunction{Trading}{Trading}
    \SetKwProg{Fn}{Function}{:}{}  
    
    \For{$i=1,2,\ldots,T$}{
        \Fn{\Signature{$X_i^{(1)}$, $X_i^{(2)}$, $\alpha$}}{
            \tcp{\emph{Calculation of Signature}}
            \KwRet $C_i^{(1,2)}$ (segmented signature), $D_i^{(1)}$ (difference of $X_i^{(1)}$), $D_i^{(2)}$ (difference of $X_i^{(2)}$)
        }
        
        \eIf{current $C_i^{(1,2)}$ < historical mean $C^{(1,2)}$}{
            \If{$D_i^{(1)} \times D_i^{(2)} > 0$ and Pair trading condition triggered}{
                signal = (short $X^{(1)}$, long $X^{(2)}$)
            }
            \ElseIf{$D_i^{(1)} \times D_i^{(2)} > 0$ and Pair trading condition triggered}{
                signal = (long $X^{(1)}$, short $X^{(2)}$)
            }
            \Else{
                signal = (hold, hold)
            }
        }{
            signal = (hold, hold)
        }
    }
    
    \Fn{\Trading{signal, $\alpha$}}{  
        \tcp{\emph{Complete the trading and calculate the results}}
        \KwRet overall return, mean daily return, max drawdown, standard deviation, Sharpe ratio, count
    }
\end{algorithm}

The condition "$current \ C_i^{(1,2)} < historical \ mean \ C^{(1,2)}$" means that the segmented signature has undergone a certain degree of change. According to the construction of segmented signature, this indicates that the interactivity or correlation of the path has become stronger. Moreover, the condition "$D_i^{(1)} \times D_i^{(2)} >0$" means that the futures increase or decrease simultaneously at the current window ($D_i^{(1,2)}>0$). Only satisfying the two conditions above at the same time, we will consider verifying whether the pair trading condition is triggered and tend to implement the trading.

\section{Application of segmented signature in pair trading}

\subsection{Calculation of segmented signature}
Given that pair trading relies on the inherent correlation between paired assets, and to ensure the robustness of our results, we categorize the futures contracts into five groups for backtesting. The basis for grouping follows the realistic correlation between different futures. According to the theory in Section 2, we introduce the calculation method of segmented signature. Calculating the signature requires a path of a certain length, in other words, it requires two time series within a window. We take the window size $w$ as 60, which means that we use the first 60 data from the starting time spot and roll the time window to calculate signature and segmented signature. When calculating the original signature, we simply discrete integral to get the final value. In order to ensure that the price data possess a certain degree of stability, we take logarithm of the price. Calculating segmented signature may be a lot bit complex, the steps are as follows:
\begin{enumerate}
    \item Preprocess: Taking log of the price data. 
    \item Interpolation: Connecting points using linear interpolation.
    \item Segmentation: Calculating the crossing points between the trajectory of the paths and its chord.
    \item Area accumulation: Calculating every enclosed area between the path and its chord between crossing points in sequence, then summing them up. 
\end{enumerate}

\begin{figure}[H]
    \centering
    \includegraphics[width=1\textwidth]{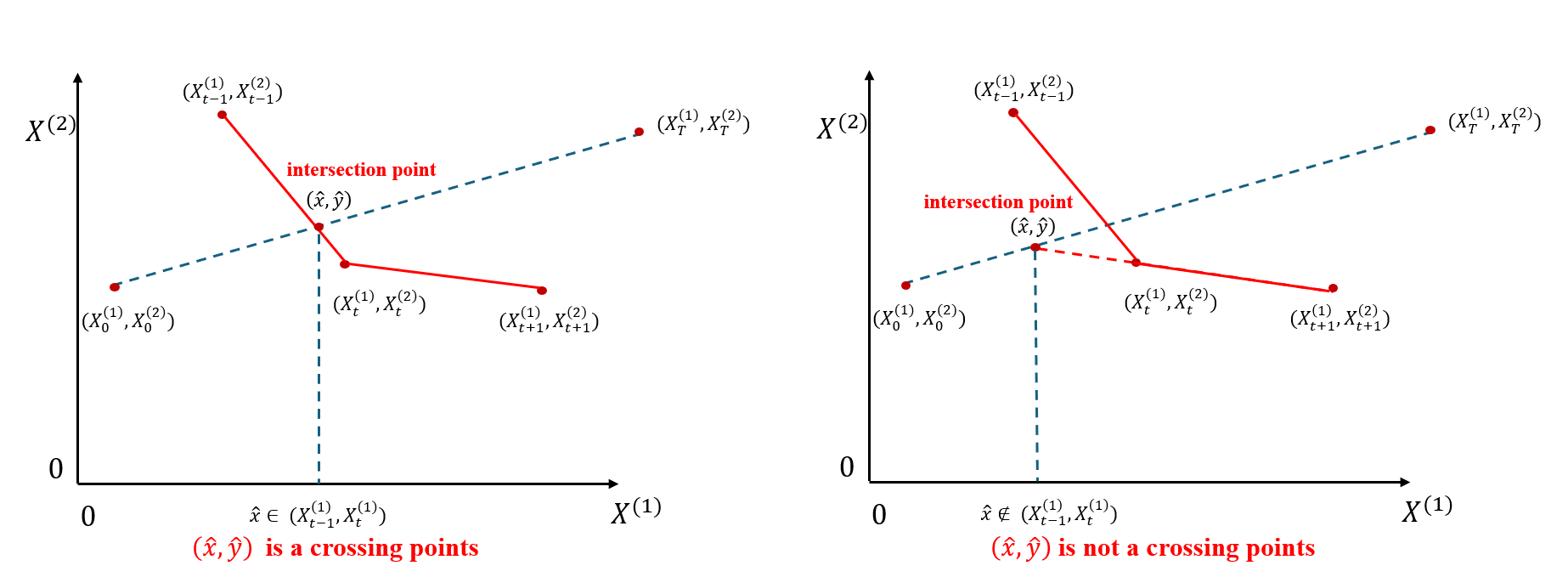}
    \caption{Identify as a Crossing Point (left) and not a Crossing Point (right)}
    \label{cross}
\end{figure}

It should be specifically noted that the calculation of crossing points in Step 3. Because we use linear interpolation to generate a continuous function from a series of discrete high-frequency data, it is easy to calculate the crossing points. Treat $X^1$ as the x-axis and $X^2$ as the y-axis. Given the starting point $(X^{(1)}_0,X^{(2)}_0)$ and the ending point $(X^{(1)}_T,X^{(2)}_T)$, we can easily derive the equation satisfied by the chord. With two sets of asset pair time series $X_0^{(1)} \cdots X_T^{(1)}$, $X_0^{(2)} \cdots X_T^{(2)}$ provided and under the linear interpolation assumption, we obtain a linear function between $(X^{(1)}_{t-1},X^{(2)}_{t-1})$ and $(X^{(1)}_{t},X^{(2)}_{t})$, then we want to test whether the crossing point is between these two points. So we calculate the intersection point between this linear function and the chord, if the x coordinate of the intersection falls between $X^{(1)}_{t-1}$ and $X^{(1)}_{t}$, then it is considered as a crossing point. To facilitate a more vivid understanding, we will illustrate it using Figure \ref{cross}.

\begin{figure}[H]
    \centering
    \includegraphics[width=0.9\textwidth]{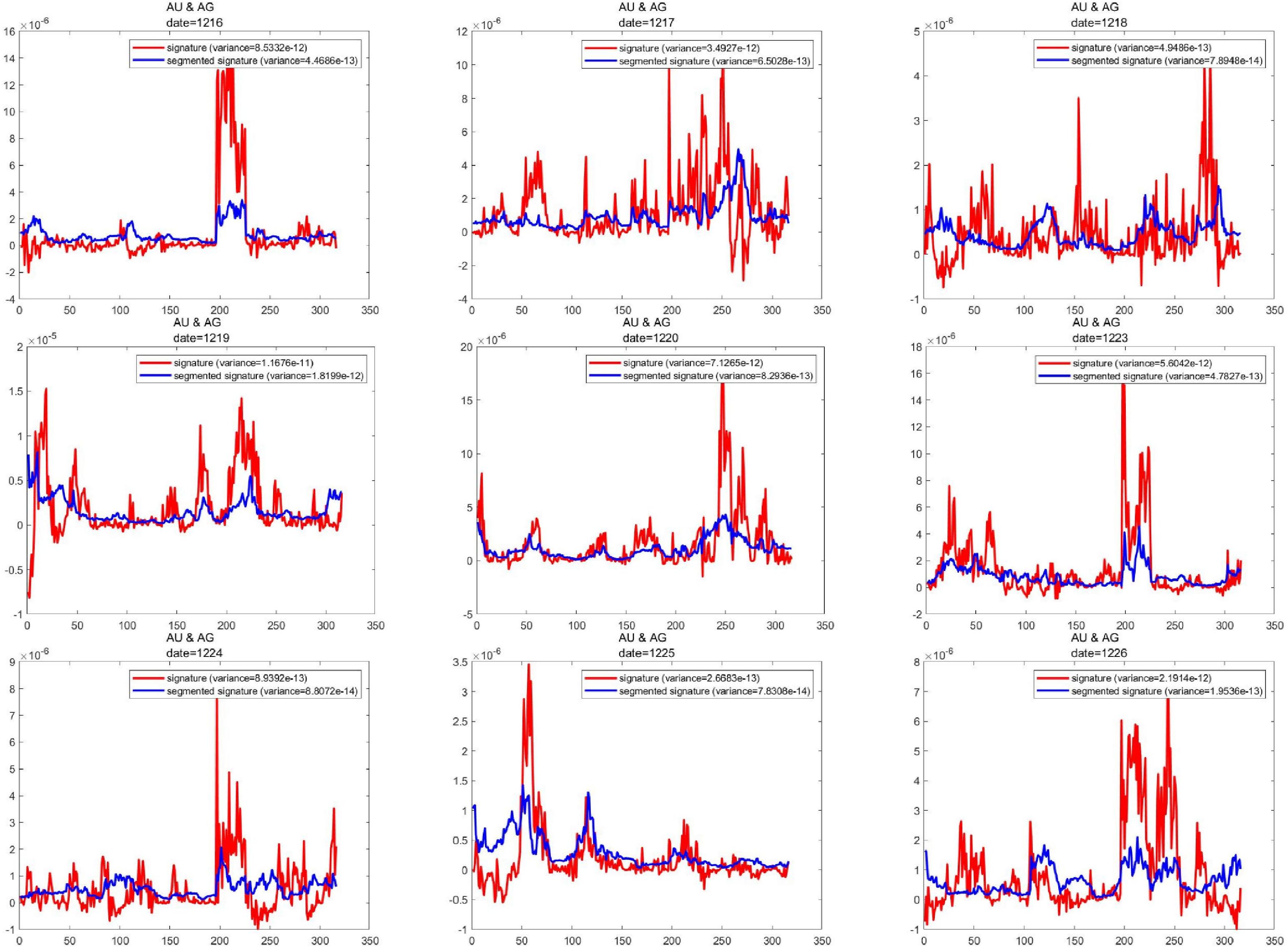}  
    \caption{Signature and Segmented Signature of AU and AG}
    \label{sta_auag}
\end{figure}

After this process, we use each crossing point as a time division to segmentally calculate the area of the polygon formed by the interpolated path and chord, and sum these areas to obtain our indicator. This corresponds to the task performed in the step 4. Taking Group 1 futures as an example, we show the daily signature and segmented signature for nine trading days for asset pair AU \& AG in Figure \ref{sta_auag}, and then analyse the relative volatility of signature and segmented signature, the results are presented in Table \ref{bianyi1}.

From the above figures, we can find that although the values of signature and segmented signature are nearly on the same order of magnitude, the values of signature are larger than those of segmented signature.  This discrepancy leads to the perception that signature exhibited greater fluctuations and variance. After conducting the ADF test, we found that both segmented signature and signature exhibit stability under statistical significance. Therefore, in order to assess the volatility and dispersion, we calculate the coefficient of variation, defined as $\frac{\sigma}{\mu}$, where $\sigma$ is the standard deviation  and $\mu$ is the mean value . Because the original signature has positive and negative signs, when calculating the coefficient of variation, we add the minimum value to all its numbers. This ensures that all values are positive and avoids the situation where the positive and negative signs cancel each other out, resulting in a very small average value $\mu$. The coefficient of variations of Group 1 assets are shown in Table \ref{bianyi1}.

\begin{table}[H]
\centering
\caption{Coefficient of variation of signature and segmented signature (Group 1)}
\resizebox{\textwidth}{!}{
\begin{tabular}{cccccccc}
\toprule 
Date&  Coefficient of variation& AU\&AG & AU\&AL & AU\&SN & AL\&AG & AG\&SN & AL\&SN\\
\midrule 
\multirow{2}{*}{1216} & signature & 0.9761&0.1568& 0.4233& 0.2618& 0.9113&0.5147\\
                     & segmented signature & 1.0484&1.2394& 1.6& 0.9649& 1.112&0.823\\ 
                    
\hline
\multirow{2}{*}{1217} & signature & 0.4551&0.3039& 0.3565& 0.221& 0.4012&0.2277\\
                     & segmented signature & 0.9486&0.8058& 1.0573& 0.7017& 0.5864&1.1364\\ 
                    
\hline
\multirow{2}{*}{1218} & signature & 0.5967&0.4239& 0.5873& 0.4551& 0.1384&0.4595\\
                     & segmented signature & 0.7798&0.678& 1.2638& 0.6972& 0.9299&0.7848\\ 
                    
\hline
\multirow{2}{*}{1219} & signature & 0.3378&0.3515& 0.1944& 0.3694& 0.2976&0.4588\\
                     & segmented signature & 0.965&0.917& 1.1333& 0.9902& 1.0345&0.9561\\ 
                    
\hline
\multirow{2}{*}{1220} & signature & 0.8652&0.2412& 0.4067& 0.4569& 0.2284&0.276\\
                     & segmented signature & 0.9046&0.8502& 0.7795& 0.7196& 0.8656&0.7558\\ 
                    
\hline
\multirow{2}{*}{1223} & signature & 1.0687&0.1113& 1.0861& 0.2552& 0.5895&0.3378\\
                     & segmented signature & 0.9385&1.2695& 1.1611& 0.7458& 0.9164&0.8435\\ 
                    
\hline
\multirow{2}{*}{1224} & signature & 0.6386&0.1301& 0.1479& 0.1186& 0.1952&0.6151\\
                     & segmented signature & 0.7527&0.954& 0.642& 1.0702& 0.7573&0.7281\\ 
                    
\hline
\multirow{2}{*}{1225} & signature & 0.7191&0.4727& 0.5267& 0.433& 0.5907&0.3354\\
                     & segmented signature & 1.0393&0.7811& 1.1238& 0.8625& 1.1065&0.6852\\ 
                    
\hline
\multirow{2}{*}{1226} & signature & 0.7972&0.5775& 0.5079& 0.8168& 0.6642&0.6624\\
                     & segmented signature & 0.8033&1.0187& 1.061& 0.9944& 1.1244&0.8768\\ 
                    
\hline
\end{tabular}}\label{bianyi1}
\end{table}

From Table \ref{bianyi1}, we find that the coefficient of variation or path fluctuation of segmented signature is generally larger that of signature. This indicates that at the same numerical level, segmented signature is relatively discrete and deviate slightly. According to this phenomenon, we think that the decomposition of signature leads to more critical information being highlighted, which means that segmented signature is more suitable to be considered as a filter because its signals are more pronounced.

\subsection{Empirical results}
In this section, we present some empirical results of different strategies and show the advantages of applying segmented signature and path difference into the strategy. Now we briefly introduce the measurement indexes for different strategies: Overall return rate: Net profit divided by initial balance; Mean daily return: Conversion of overall return rate to daily return rate; Max drawdown: Maximum drawdown; Std: Standard deviation; Sharpe ratio: Sharpe ratio calculated by excess returns; Count: Number of transactions. The empirical results are shown as follows.

\begin{table}[H]
\centering
\caption{Performance of Different Strategies on Futures Pairs (Group 1)}
\resizebox{\textwidth}{!}{
\begin{tabular}{cccccccc}
\toprule 
futures &   Strategies &Overall return rate& Mean daily return & Max drawdown & Std & Sharpe ratio &Count\\
\midrule 
\multirow{4}{*}{AU\&AG} & NO SIG & 2.27\% &0.041\% & -1.95\% & 0.55 & 1.00 &2398\\
                     & SIG & -0.13\% &-0.0039\% & -2.19\% & 0.56 & -0.28 &1647\\ 
                     & SE-SIG & 2.13\% &0.039\% & -1.61\% & 0.41 & 1.29 &1633\\
                     & SE-SIG-DIFF & 2.64\% &0.048\% & -1.29\% & 0.46 & 1.44 &1335\\ 
\hline
\multirow{4}{*}{AU\&AL} & NO SIG &2.48\%&0.045\%&-1.64\%&0.40&1.57&1801\\
                     & SIG &0.66\%&0.011\%&-2.92\%&0.45&0.19&1169\\ 
                     & SE-SIG &3.79\%&0.069\%&-1.74\%&0.41&2.48&1195\\
                     & SE-SIG-DIFF &3.74\%&0.069\%&-1.57\%&0.35&2.83&667\\ 
\hline
\multirow{4}{*}{AU\&SN} & NO SIG &-3.95\%&-0.078\%&-7.31\%&0.62&-2.14&1672\\
                     & SIG  &0.13\%&0.0003\%&-4.61\%&0.64&-0.14&1094\\
                     & SE-SIG &0.32\%&0.0046\%&-3.14\%&0.55&-0.04&1168\\
                     & SE-SIG-DIFF &3.94\%&0.072\%&-1.41\%&0.50&2.10&778\\
\hline
\multirow{4}{*}{AL\&AG} & NO SIG &1.42\%&0.024\%&-2.99\%&0.65&0.45&1783\\ 
                     & SIG &1.31\%&0.023\%&-2.40\%&0.54&0.50&1266\\ 
                     & SE-SIG &3.03\%&0.054\%&-2.76\%&0.62&1.23&1226\\
                     & SE-SIG-DIFF &6.63\%&0.12\%&-1.81\%&0.59&3.03&550\\
\hline
\multirow{4}{*}{AG\&SN} & NO SIG &-3.92\%&-0.078\%&-8.60\%&0.74&-1.81&1670\\ 
                     & SIG &-0.44\%&-0.01\%&-5.64\%&0.64&-0.40&1069\\
                     & SE-SIG &2.54\%&0.045\%&-4.13\%&0.66&0.95&1140\\
                     & SE-SIG-DIFF &2.75\%&0.049\%&-4.58\%&0.64&1.06&832\\ 
\hline
\multirow{4}{*}{AL\&SN} & NO SIG &-1.62\%&-0.033\%&-6.05\%&0.65&-0.95&1664\\ 
                     & SIG &0.30\%&0.0037\%&-4.32\%&0.63&-0.06&958\\
                     & SE-SIG &2.93\%&0.053\%&-4.54\%&0.61&1.21&1096\\ 
                     & SE-SIG-DIFF &2.57\%&0.046\%&-3.38\%&0.63&1.01&742\\
\hline
\end{tabular}}\label{Tableg1}
\end{table}
Table \ref{Tableg1} presents the performance of each strategy on different futures in Group 1, which is the metal futures. The results show a significant increase in return (both the overall return and the mean daily return) with using segmented signature (SE-SIG) and the segmented signature and price difference product (SE-SIG-DIFF) as filtering signal, compared to pair trading with no filtering signal (NO SIG) and with original signature (SE-SIG). Also, surprisingly, the max drawdown has a significant decrease after the filtering with segmented signature and price difference product (SE-SIG-DIFF) signal, which indicates a simultaneous improvement on increasing return and decreasing risk. These results show the strong potential of the SE-SIG-DIFF strategy in arbitrage models.  

\begin{table}[H]
\centering
\caption{Comparison of Sharpe Ratio of Different Strategies (Group 1)} 
\resizebox{0.7\textwidth}{!}{
\begin{tabular}{lccccll}
\toprule 
Sharpe ratio & AU\&AG & AU\&AL  
& AU\&SN& AL\&AG &  AG\&SN  &AL\&SN\\
\midrule 
NO SIG & 1& 1.57& -2.14& 0.45&  -1.81&-0.95\\
SIG& -0.28& 0.19& -0.14& 0.50&  -0.40&-0.06\\
SE-SIG& 1.29& 2.48& -0.04& 1.23&  0.95&\bm{\textcolor{red}{1.21}}\\
SE-SIG-DIFF& \bm{\textcolor{red}{1.44}} & \bm{\textcolor{red}{2.83}} & \bm{\textcolor{red}{2.10}} & \bm{\textcolor{red}{3.03}} & \bm{\textcolor{red}{1.06}}&1.01\\
\bottomrule 
\end{tabular}} \label{sharpeg1}
\end{table}

Table \ref{sharpeg1} illustrates the comparison of Sharpe ratio of different methods. The results reveal that the Sharpe ratio has significantly increased after using segmented signature and path difference product (SE-SIG-DIFF) as filters. Although in $\text{AL\&SN}$ pair of assets, SE-SIG strategy produced a little bit more profit than SE-SIG-DIFF strategy, but the max drawdown of SE-SIG-DIFF is lower than SE-SIG, which show less risk of SE-SIG-DIFF. Overall, SE-SIG-DIFF shows the greater profitability, lower risk, and more robust performance.

\begin{figure}[H]
    \centering
    \includegraphics[width=1\textwidth]{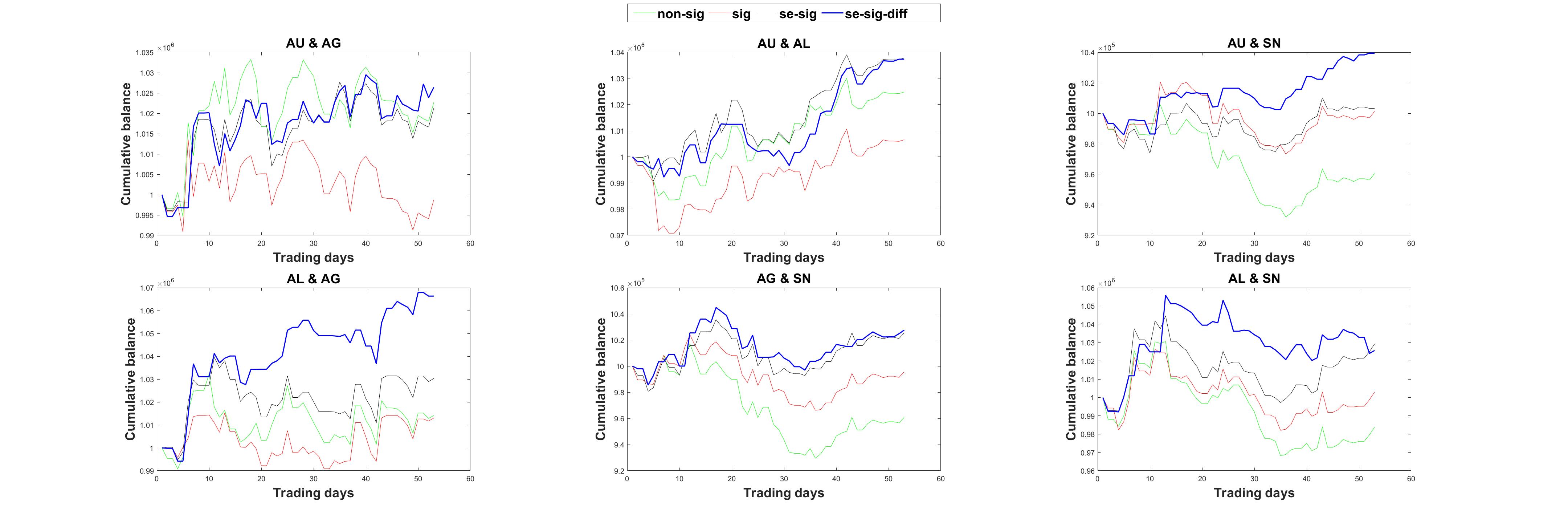}  
    \caption{Cumulative Balance of Different Strategies (Group 1)}\label{cumug1}
\end{figure}

Figure \ref{cumug1} illustrates the comparison of the cumulative balance of different strategies in Group 1. The figures exhibit that the SE-SIG-DIFF strategy performs better than other strategies, since it outperforms other strategies at most of the time. And from the perspective of profit and risk, SE-SIG-DIFF strategy generates greater profits during the period when all strategies generate profits, and generates smaller losses during the period when all strategies generate losses. Additionally, compared with other strategies, SE-SIG-DIFF strategy is capable of turning the losses into gains while other strategies incur losses.

\begin{table}[H]
\centering
\caption{Performance of Different Strategies on Futures Pairs (Group 2)}
\resizebox{\textwidth}{!}{
\begin{tabular}{cccccccc}
\toprule 
futures &   Strategies &Overall return rate& Mean daily return & Max drawdown & Std & Sharpe ratio &Count\\
\midrule 
\multirow{4}{*}{C\&B} & NO SIG &3.40\%&0.074\%&-1.87\%&0.57&1.89&5077\\
                     & SIG &3.33\%&0.073\%&-1.80\%&0.57&1.85&3373\\ 
                     & SE-SIG &3.33\%&0.073\%&-1.80\%&0.57&1.85&3197\\
                     & SE-SIG-DIFF &4.42\%&0.096\%&-2.35\%&0.59&2.42&1712\\
\hline
\multirow{4}{*}{C\&CF} & NO SIG &-2.99\%&-0.07\%&-4.45\%&0.49&-2.48&4592\\ 
                     & SIG &-3.42\%&-0.08\%&-4.46\%&0.46&-2.94&3122\\
                     & SE-SIG &-3.33\%&-0.078\%&-5.21\%&0.50&-2.67&2932\\
                     & SE-SIG-DIFF &-0.93\%&-0.022\%&-2.36\%&0.45&-0.99&1349\\
\hline
\multirow{4}{*}{C\&M} & NO SIG &-1.59\%&-0.038\%&-2.46\%&0.50&-1.39&4743\\ 
                     & SIG &-2.60\%&-0.061\%&-3.44\%&0.50&-2.13&3100\\
                     & SE-SIG &-2.20\%&-0.052\%&-2.55\%&0.51&-1.80&3014\\
                     & SE-SIG-DIFF &-0.89\%&-0.022\%&-2.44\%&0.52&-0.84&1633\\
\hline
\multirow{4}{*}{B\&CF} & NO SIG &-7.45\%&-0.18\%&-8.45\%&0.59&-4.92&4736\\
                     & SIG &-8.40\%&-0.2\%&-9.38\%&0.60&-5.49&3240\\
                     & SE-SIG &-6.15\%&-0.15\%&-6.79\%&0.60&-4.00&3037\\
                     & SE-SIG-DIFF &-4.79\%&-0.11\%&-6.12\%&0.62&-3.07&1414\\
\hline
\multirow{4}{*}{B\&M} & NO SIG &4.78\%&0.1\%&-3.05\%&0.56&2.80&4890\\
                     & SIG &4.57\%&0.099\%&-3.75\%&0.63&2.36&3539\\ 
                     & SE-SIG &5.82\%&0.13\%&-1.78\%&0.58&3.30&3292\\ 
                     & SE-SIG-DIFF &6.64\%&0.14\%&-1.80\%&0.60&3.66&2252\\ 
\hline
\multirow{4}{*}{M\&CF} & NO SIG &-7.65\%&-0.18\%&-7.91\%&0.60&-4.96&4755 \\
                     & SIG &-6.26\%&-0.15\%&-6.85\%&0.54&-4.50&3271\\
                     & SE-SIG &-7.17\%&-0.17\%&-7.17\%&0.60&-4.68&3207 \\
                     & SE-SIG-DIFF &-5.59\%&-0.13\%&-6.32\%&0.57&-3.86&1542 \\
\hline
\end{tabular}}\label{Tableg2}
\end{table}

\begin{table}[H]
\centering
\caption{Comparison of Sharpe Ratio of Different Strategies (Group 2)} 
\resizebox{0.7 \textwidth}{!}{
\begin{tabular}{lccccll}
\toprule 
Sharpe ratio & C\&B & C\&CF  
& C\&M & B\&CF &  B\&M  & M\&CF \\
\midrule 
NO SIG & 1.89& -2.48& -1.39& -4.92&  2.80 &-4.96\\
SIG & 1.85& -2.94& -2.13& -5.49& 2.36&-4.50\\
SE-SIG& 1.85& -2.67& -1.80& -4.00& 3.30 &-4.68\\
SE-SIG-DIFF& \bm{\textcolor{red}{2.42}} & \bm{\textcolor{red}{-0.99}} & \bm{\textcolor{red}{-0.84}}& \bm{\textcolor{red}{-3.07}}& \bm{\textcolor{red}{3.66}} & \bm{\textcolor{red}{-3.86}}\\

\bottomrule 
\end{tabular}}\label{sharpeg2}
\end{table}

Table \ref{Tableg2} and \ref{sharpeg2} reveal the performance of each strategy on different futures and compare the Sharpe ratio of different methods in Group 2 (agricultural product futures). The results also indicate the strong evidence that SE-SIG-DIFF is able to increase return and reduce risk (which is measured by max drawdown and std) in most cases. The Sharpe ratio in Table \ref{sharpeg2} exhibits that the SE-SIG-DIFF strategy performs better in all pairs of assets. When significant losses or systematic risks arise, the SE-SIG-DIFF strategy is capable of helping control certain risks. While traditional strategies are effective, the SE-SIG-DIFF strategy is able to increase returns.

\begin{figure}[H]
    \centering
    \includegraphics[width=1\textwidth]{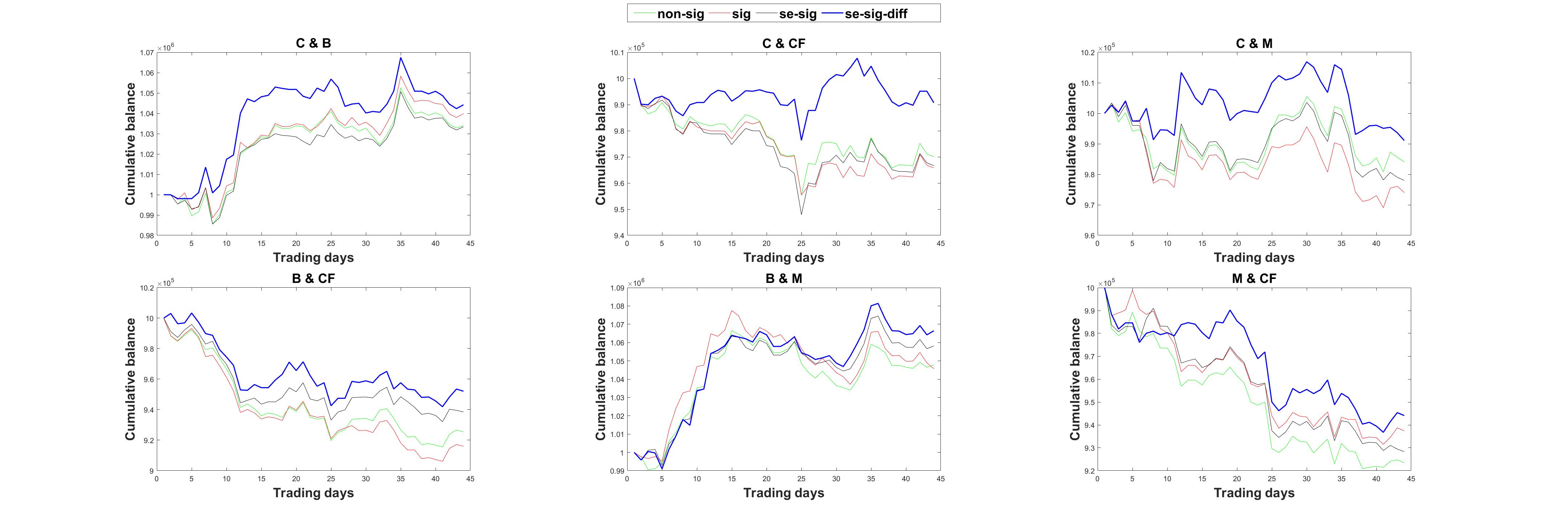}  
    \caption{Cumulative Balance of Different Strategies (Group 2)}\label{cumug2}
\end{figure}

Figure \ref{cumug2} shows the comparison of the cumulative balance of different strategies in Group 2. The figure clearly illustrates that the SE-SIG-DIFF strategy is able to outperform other strategies and gain more profit with the higher cumulative balance compared with other strategies.

\begin{table}[H]
\centering
\caption{Performance of Different Strategies on Futures Pairs (Group 3)}
\resizebox{\textwidth}{!}{
\begin{tabular}{ccccccccc}
\toprule 
futures &   Strategies &Overall return & Mean daily return & Max drawdown & Std & Sharpe ratio &Count\\
\midrule 
\multirow{4}{*}{MA\&SC} & NO SIG &0.57\%&0.012\%&-2.67\%&0.49&0.19&2491\\
                     & SIG &0.51\%&0.01\%&-2.16\%&0.46&0.15&1605\\ 
                     & SE-SIG &0.39\%&0.0076\%&-2.90\%&0.50&0.05&1686\\
                     & SE-SIG-DIFF &3.54\%&0.078\%&-1.95\%&0.50&2.30&866\\
\hline
\multirow{4}{*}{MA\&Y} & NO SIG &-2.82\%&-0.067\%&-5.44\%&0.63&-1.85&2870\\
                     & SIG &-3.57\%&-0.085\%&-5.76\%&0.63&-2.27&1746\\
                     & SE-SIG &-3.05\%&-0.072\%&-5.70\%&0.62&-1.99&1747\\
                     & SE-SIG-DIFF  &-2.47\%&-0.059\%&-6.96\%&0.72&-1.45&794\\
\hline
\multirow{4}{*}{MA\&RB} & NO SIG &1.29\%&0.028\%&-2.75\%&0.49&0.71&2686\\
                     & SIG &1.50\%&0.033\%&-2.74\%&0.46&0.92&1843\\
                     
                     & SE-SIG &2.96\%&0.065\%&-1.87\%&0.47&1.98&1819\\
                     & SE-SIG-DIFF &2.22\%&0.049\%&-1.71\%&0.47&1.47&928\\
\hline
\multirow{4}{*}{SC\&Y} & NO SIG &5.89\%&0.13\%&-2.21\%&0.69&2.82&2499 \\
                     & SIG &6.32\%&0.14\%&-2.83\%&0.64&3.28&1576 \\
                     & SE-SIG &9.91\%&0.21\%&-2.17\%&0.64&5.09&1590\\
                     & SE-SIG-DIFF &10.14\%&0.22\%&-1.63\%&0.62&5.38&813 \\
\hline
\multirow{4}{*}{SC\&RB} & NO SIG &3.54\%&0.078\%&-1.33\%&0.42&2.72&4646 \\ 
                     & SIG &2.25\%&0.05\%&-2.80\%&0.46&1.52&3175 \\ 
                     & SE-SIG &1.00\%&0.021\%&-3.53\%&0.54&0.45&3301 \\
                     & SE-SIG-DIFF &5.40\%&0.12\%&-1.08\%&0.45&3.95&1669 \\ 
\hline
\multirow{4}{*}{RB\&Y} & NO SIG &-2.09\%&-0.05\%&-5.62\%&0.57&-1.55&5122 \\
                     & SIG &-1.95\%&-0.046\%&-5.29\%&0.60&-1.38&3587 \\
                     & SE-SIG &-0.59\%&-0.015\%&-5.08\%&0.57&-0.59&3610 \\ 
                     & SE-SIG-DIFF &0.53\%&0.01\%&-4.76\%&0.60&0.12&1617 \\ 
\hline
\end{tabular}}\label{Tableg3}
\end{table}

\begin{table}[H]
\centering
\caption{Comparison of Sharpe Ratio of Different Strategies (Group 3)} 
\resizebox{0.7\textwidth}{!}{
\begin{tabular}{lccccll}
\toprule 
Sharpe ratio & MA\&SC & MA\&Y  
& MA\&RB& SC\&Y &  SC\&RB  &RB\&Y\\
\midrule 
NO SIG & 0.19& -1.85& 0.92& 2.82&  2.72&-1.55\\
SIG& 0.15& -2.27& 0.71& 3.28& 1.52&-1.38\\
SE-SIG& 0.05& -1.99& \bm{\textcolor{red}{1.98}}& 5.09& 0.45&-0.59\\
SE-SIG-DIFF& \bm{\textcolor{red}{2.30}} & \bm{\textcolor{red}{-1.45}} & 1.47& \bm{\textcolor{red}{5.38}}& \bm{\textcolor{red}{3.86}} & \bm{\textcolor{red}{0.12}}\\

\bottomrule 
\end{tabular}}\label{sharpeg3}
\end{table}

Similarly, the Table \ref{Tableg3} and \ref{sharpeg3} show the the performance of each strategy on different futures and the comparison of Sharpe ratio of different methods in Group 3 (Oil related products futures). The results indicate some negative impact of original signature on the transaction. And the advantages of SE-SIG and SE-SIG-DIFF are gradually reflected, especially SE-SIG-DIFF, which has significant role on increasing returns, improving Sharpe ratio, and reducing max drawdown.

\begin{figure}[H]
    \centering
    \includegraphics[width=1\textwidth]{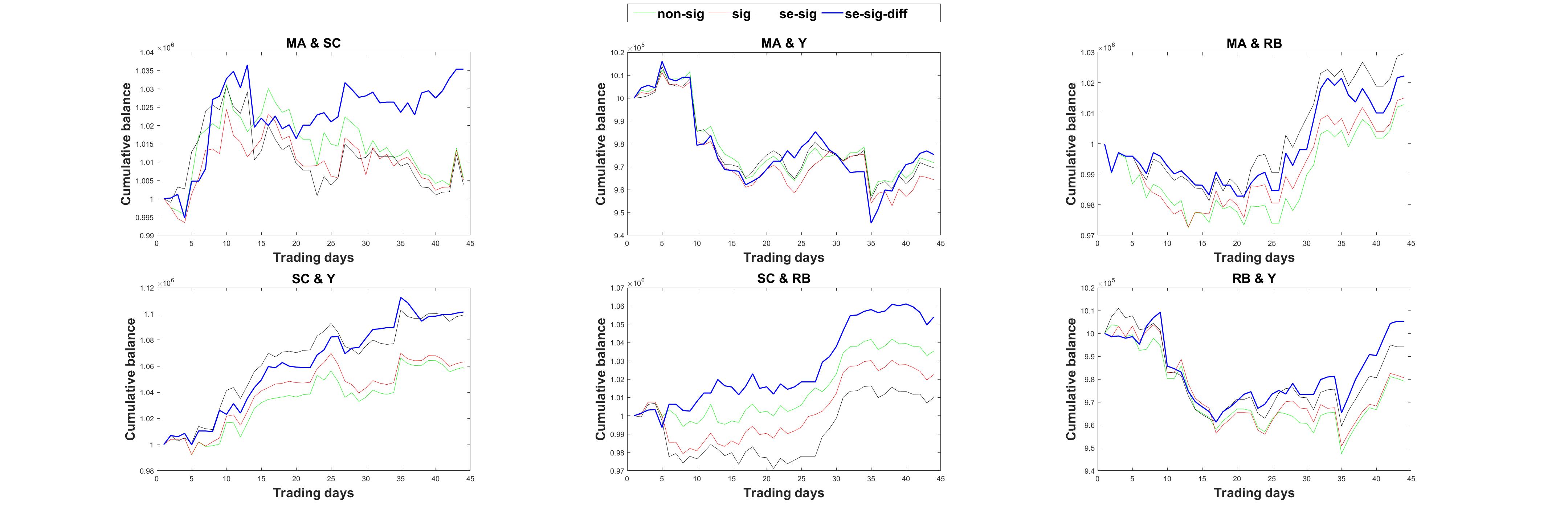}  
    \caption{Cumulative Balance of Different Strategies (Group 3)}\label{cumug3}
\end{figure}

Also, the comparison of the cumulative balance of different strategies in Group 3 is shown in Figure \ref{cumug3}. The figure presents a relatively strong ability to make profit from SE-SIG and the SE-SIG-DIFF strategy with the leading performance of the cumulative balance.

\begin{figure}[H]
    \centering
    \includegraphics[width=1\textwidth]{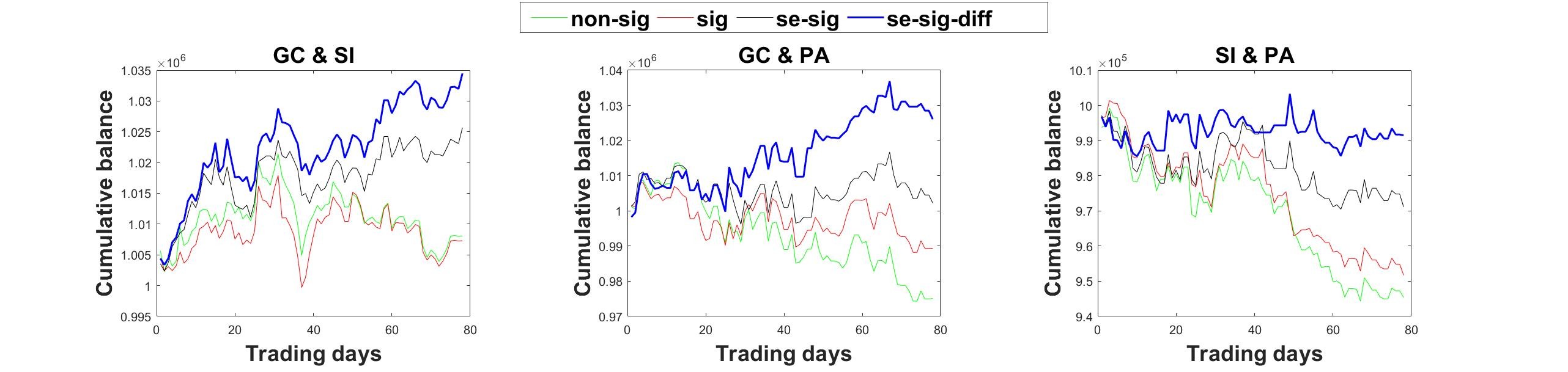}  
    \caption{Cumulative Balance of Different Strategies (Group 4)}\label{cumug4}
\end{figure}

\begin{figure}[H]
    \centering
    \includegraphics[width=1\textwidth]{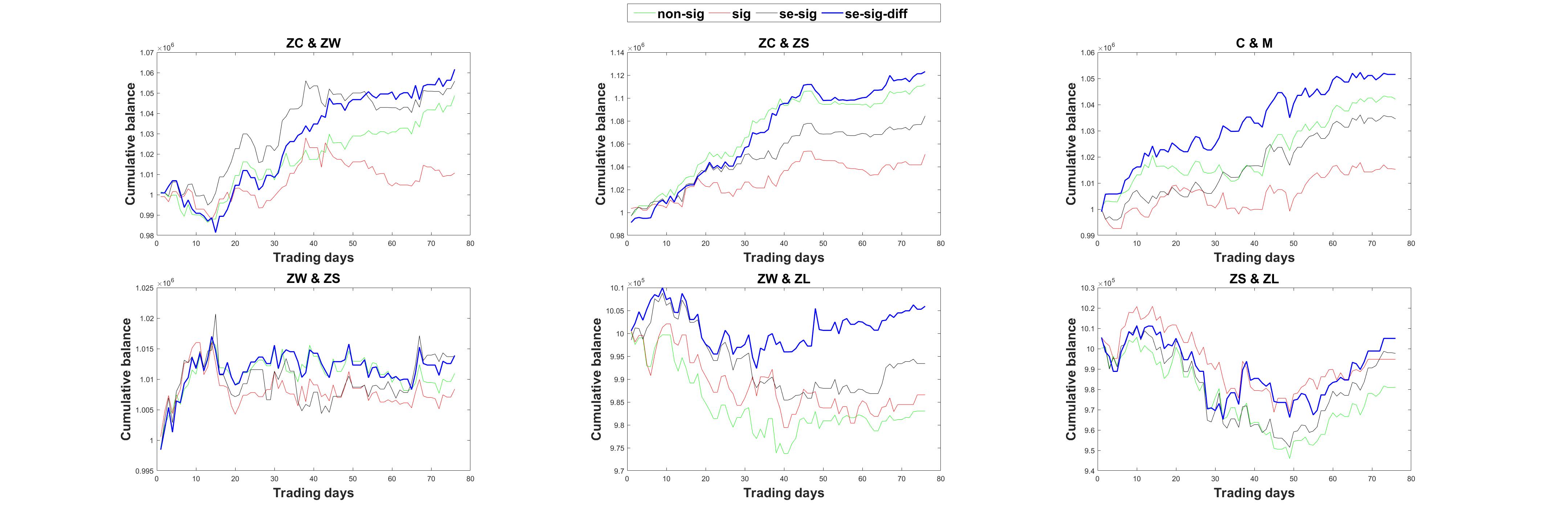}  
    \caption{Cumulative Balance of Different Strategies (Group 5)}\label{cumug5}
\end{figure}

For Groups 4 and 5, the results are similarly presented (Group 4: Table \ref{Tableg4}, Table \ref{sharpeg4}, Figure \ref{cumug4}; Group 5: Table \ref{Tableg5}, Table \ref{sharpeg5}, Figure \ref{cumug5}). We observed that the findings are consistent with those of first three groups. Specifically, the SE-SIG strategy provided a certain degree of improvement over the baseline strategy, underscoring the effectiveness of the segmented signature $C_{s,t}^{(1,2)}$. However, the SE-SIG-DIFF strategy demonstrated superior performance, which validates our initial hypothesis and confirms that both the $C_{s,t}^{(1,2)}$ and $D_{s,t}^{(1,2)}$ are critical indicators.

\begin{table}[H]
\centering
\caption{Performance of Different Strategies on Futures Pairs (Group 4)}
\resizebox{\textwidth}{!}{
\begin{tabular}{ccccccccc}
\toprule 
futures &  Strategies &Overall return & Mean daily return & Max drawdown & Std & Sharpe ratio &Count\\
\midrule 
\multirow{4}{*}{GC\&SI} & NO SIG &0.81\%&0.01\%&-1.71\%&0.22&0.30& 10011 \\
                     & SIG &0.73\%&0.0091\%&-1.79\%&0.22&0.22&6598 \\ 
                     & SE-SIG &2.57\%&0.032\%&-1.01\%&0.20&2.12& 6769\\
                     & SE-SIG-DIFF &3.45\%&0.043\%&-1.05\%&0.20&2.91& 5722\\
                    
\hline
\multirow{4}{*}{GC\&PA} & NO SIG &-2.49\%&-0.033\%&-3.88\%&0.34&-1.81&4257 \\
                     & SIG &-1.06\%&-0.014\%&-1.97\%&0.32&-1.00&3018 \\
                     & SE-SIG &0.22\%&0.0023\%&-1.65\%&0.33&-0.17&2675 \\
                     & SE-SIG-DIFF  &2.61\%&0.032\%&-1.15\%&0.35&1.21&1289 \\
\hline
\multirow{4}{*}{SI\&PA} & NO SIG &-5.46\%&-0.073\%&-5.49\%&0.36&-3.42&4318 \\
                     & SIG &-4.83\%&-0.064\%&-4.96\%&0.33&-3.32&2592 \\
                     
                     & SE-SIG &-2.89\%&-0.038\%&-2.79\%&0.40&-1.78&2789 \\
                     & SE-SIG-DIFF &-0.85\%&-0.012\%&-1.76\%&0.33&-0.84&1969 \\
\hline
\end{tabular}}\label{Tableg4}
\end{table}

\begin{table}[H]
\centering
\caption{Comparison of Sharpe Ratio of Different Strategies (Group 4)} 
\resizebox{0.5\textwidth}{!}{
\begin{tabular}{lcll}
\toprule 
Sharpe ratio & GC\&SI & GC\&PA  
& SI\&PA\\
\midrule 
NO SIG & 0.30& -1.70& -3.42\\
SIG& -1.81& -1.00& -3.32\\
SE-SIG& 2.12& -0.17& -1.78\\
SE-SIG-DIFF& \bm{\textcolor{red}{2.91}} & \bm{\textcolor{red}{1.21}} &\bm{\textcolor{red}{-0.84}}\\
\bottomrule 
\end{tabular}}\label{sharpeg4}
\end{table}

\begin{table}[H]
\centering
\caption{Performance of Different Strategies on Futures Pairs (Group 5)}
\resizebox{\textwidth}{!}{
\begin{tabular}{ccccccccc}
\toprule 
futures &  Strategies &Overall return & Mean daily return & Max drawdown & Std & Sharpe ratio &Count\\
\midrule 
\multirow{4}{*}{ZC\&ZW} & NO SIG &4.88\%&0.062\%&-2.53\%&0.33&2.67&4131 \\
                     & SIG &1.06\%&0.013\%&-2.30\%&0.36&0.33&2841\\ 
                     & SE-SIG &5.59\%&0.071\%&-1.48\%&0.39&2.67&2571 \\
                     & SE-SIG-DIFF &6.16\%&0.078\%&-1.50\%&0.35&3.25& 1394 \\
\hline
\multirow{4}{*}{ZC\&ZS} & NO SIG&11.21\%&0.14\%&-1.31\%&0.37&5.68&4504 \\
                     & SIG &5.09\%&0.064\%&-2.02\%&0.40&2.31& 3030\\
                     & SE-SIG &8.44\%&0.11\%&-1.11\%&0.38&4.14&2892 \\
                     & SE-SIG-DIFF  &12.31\%&0.15\%&-1.26\%&0.39&5.92&1681 \\
\hline
\multirow{4}{*}{ZC\&ZL} & NO SIG &4.20\%&0.054\%&-0.96\%&0.21&3.65&3980 \\
                     & SIG &1.52\%&0.02\%&-1.10\%&0.23&0.92& 2266\\
                     
                     & SE-SIG &3.45\%&0.044\%&-0.92\%&0.20&2.99&2511 \\
                     & SE-SIG-DIFF &5.15\%&0.066\%&-0.78\%&0.24&3.94& 1417\\
\hline
\multirow{4}{*}{ZW\&ZS} & NO SIG &1.09\%&0.014\%&-0.84\%&0.17&0.75&4580 \\
                     & SIG &0.84\%&0.011\%&-1.16\%&0.19&0.40&3262 \\
                     & SE-SIG &1.37\%&0.018\%&-1.60\%&0.24&0.76&2939\\
                     & SE-SIG-DIFF &1.39\%&0.018\%&-0.85\%&0.20&0.95& 1350 \\
\hline
\multirow{4}{*}{ZW\&ZL} & NO SIG &-1.70\%&-0.023\%&-2.65\%&0.23&-1.99&4682 \\ 
                     & SIG &-1.34\%&-0.018\%&-2.26\%&0.24&-1.60& 2855\\ 
                     & SE-SIG &-0.66\%&-0.0089\%&-2.32\%&0.19&-1.22&2802 \\
                     & SE-SIG-DIFF &0.59\%&0.0076\%&-1.74\%&0.21&0.12& 1742\\
\hline
\multirow{4}{*}{ZS\&ZL} & NO SIG &-1.90\%&-0.027\%&-5.93\%&0.50&-1.03&5132 \\
                     & SIG &-0.52\%&-0.0081\%&-5.10\%&0.49&-0.46&3372 \\
                     & SE-SIG &-0.23\%&-0.0047\%&-5.94\%&0.56&-0.30&3106 \\ 
                     & SE-SIG-DIFF &0.50\%&0.0052\%&-4.55\%&0.53&-0.02&2627\\ 
\hline
\end{tabular}}\label{Tableg5}
\end{table}

\begin{table}[H]
\centering
\caption{Comparison of Sharpe Ratio of Different Strategies (Group 5)} 
\resizebox{0.7\textwidth}{!}{
\begin{tabular}{lccccll}
\toprule 
Sharpe ratio & ZC\&ZW & ZC\&ZS  
& ZC\&ZL& ZW\&ZS &  ZW\&ZL  &ZS\&ZL\\
\midrule 
NO SIG & 2.67& 5.68& 3.65& 0.75&  -1.99&-1.03\\
SIG& 0.33& 2.31 & 0.92& 0.40& -1.60&-0.46\\
SE-SIG& 2.67& 4.14& 2.99& 0.76& -1.22&-0.30\\
SE-SIG-DIFF& \bm{\textcolor{red}{3.25}} & \bm{\textcolor{red}{5.92 }} & \bm{\textcolor{red}{3.94}}& \bm{\textcolor{red}{0.95}}& \bm{\textcolor{red}{0.12}} & \bm{\textcolor{red}{-0.02}}\\

\bottomrule 
\end{tabular}}\label{sharpeg5}
\end{table}

In general, there are four main findings in our results. Firstly, the SE-SIG-DIFF strategy performed the best in most asset pairs when evaluated based on the Sharpe ratio. It should be noted that while the majority strategies have negative Sharpe ratios, the SE-SIG-DIFF strategy achieves a positive Sharpe ratio (Group 1: AU\&SN, AG\&SN, AL\&SN; Group 3: RB\&Y; Group 4:GC\&PA; Group 5: ZW\&ZL). What's more, in some cases, using the original signature (SIG) as a filter may enlarge the loss because it contains chaotic information mixed together, while SE-SIG-DIFF strategy performs well since it discretes the useful information (Group 1: AU\&AG, AU\&AL; Group 2: C\&CF, C\&M, B\&CF; Group 3: MA\&Y, MA\&RB, SC\&RB; Group 4: GC\&SI, GC\&PA;Group 5: ZW\&ZL).

Secondly, when we focus on overall return and standard deviation, the results reveal that the SE-SIG-DIFF strategy mainly increases the Sharpe ratio by increasing the yield, rather than reducing the standard deviation. These examples (Group 1: AL\&SN; Group 2: C\&B, C\&CF, C\&M, B\&CF; Group 3: MA\&SC, MA\&Y, MA\&RB, SC\&RB, RB\&Y; Groop 4: GC\&SI, SI\&PA; Group 5: ZC\&ZW, ZC\&ZL, ZS\&ZL) illustrate that under the SE-SIG-DIFF strategy, the standard deviation of assets increased or remained, but the yield increased more significantly, leading to an increase in Sharpe ratio.

Thirdly, for the crucial risk measurement, max drawdown, for the industry, the results present a significantly decrease on max drawdown on most pairs with the SE-SIG-DIFF strategy, which is quite useful for investors to control the risk of strategies.

Finally, the number of futures trading transactions decreased under the SE-SIG-DIFF strategy, which means that some unprofitable transactions are filtered out. This leads to the improvement in profit and also the reduction in transaction fees. The above results and the analysis of advantages all demonstrate that the trading strategy we proposed based on the decomposed signature has ideal performance.

\subsection{Robustness and statistical testing}
In this subsection, we complete the robust testing and hypothesis testing to illustrate that the improvements of strategies are not due to chance. Two parts are arranged to demonstrate this matter: 1. Robustness test; 2. Significance test.

First, we employ the robustness test by setting different conditions, and see under this kind of processing, whether the performance of the SE-SIG-DIFF strategy is still solidly advantageous. So we conduct a sensitivity analysis using various window sizes and $Z_{\text{score}}$ in pair trading strategy to verify whether our method leads to universal improvement. We set $Z_{\text{score}} = 1.5, \ 2,\  2.5$, and the window size (shorted as w in Table) as 30, 60, and 90. Then, we combine these values to calculate the Sharpe ratio under different conditions. In every group, we present the average increase in the Sharpe ratio for different strategies relative to the benchmark strategy (NO-SIG), shown in Table \ref{robust}.

\begin{table}[H]
\centering
\caption{Average Improvement of Sharpe Ratio in Different Groups} 
\resizebox{0.7\textwidth}{!}{
\begin{tabular}{lcccll}
\toprule 
$Z_{\text{score}}=1.5, w=30$ & Gruop 1 & Gruop 2 
& Gruop 3 & Gruop 4 &  Gruop 5  \\
\midrule 
SIG& 0.11& 0.46& -0.18& -1.35&  -0.34\\
SE-SIG& 1.73& 0.81& 1.29& -0.65&  0.06\\
SE-SIG-DIFF& \bm{\textcolor{red}{2.29}} & \bm{\textcolor{red}{0.95}}& \bm{\textcolor{red}{1.36}}& \bm{\textcolor{red}{0.29}}& \bm{\textcolor{red}{0.52}}\\
\bottomrule 
\addlinespace[2pt]
$Z_{\text{score}}=1.5, w=60$ & Gruop 1 & Gruop 2 
& Gruop 3 & Gruop 4 &  Gruop 5  \\
\midrule 
SIG& 0.14& 0.15& -0.09& -0.7&  -0.90\\
SE-SIG& 1.93& 0.72& 0.16& 0.25&  0.59\\
SE-SIG-DIFF& \bm{\textcolor{red}{2.72}}& \bm{\textcolor{red}{1.29}}& \bm{\textcolor{red}{1.34}}& \bm{\textcolor{red}{0.55}}& \bm{\textcolor{red}{1.41}}\\
\bottomrule 
\addlinespace[2pt]
$Z_{\text{score}}=1.5, w=90$ & Gruop 1 & Gruop 2 
& Gruop 3 & Gruop 4 &  Gruop 5  \\
\midrule 
SIG& -0.94& 0.63& 0.03& -0.29&  0.58\\
SE-SIG& -0.33& 0.86& \bm{\textcolor{red}{0.29}}& 0.36&  0.65\\
SE-SIG-DIFF& \bm{\textcolor{red}{1.09}}& \bm{\textcolor{red}{1.25}}& 0.23& \bm{\textcolor{red}{1.11}}& \bm{\textcolor{red}{0.88}}\\
\bottomrule 
\addlinespace[2pt]
$Z_{\text{score}}=2, w=30$ & Gruop 1 & Gruop 2 
& Gruop 3 & Gruop 4 &  Gruop 5 \\
\midrule 
SIG& -0.57& 0.12& 0.82& 0.42&  0.09\\
SE-SIG& 0.65& 0.23& 1.85& 0.25&  0.40\\
SE-SIG-DIFF& \bm{\textcolor{red}{1.12}} & \bm{\textcolor{red}{1.23}}& \bm{\textcolor{red}{1.87}}& \bm{\textcolor{red}{1.27}}& \bm{\textcolor{red}{1.48}}\\
\bottomrule 
\addlinespace[2pt]
$Z_{\text{score}}=2, w=60$ & Gruop 1 & Gruop 2 
& Gruop 3 & Gruop 4 &  Gruop 5 \\
\midrule 
SIG& 0.28& -0.30& -0.21& -0.44&  -1.31\\
SE-SIG& 1.50& 0.18& 0.29& 1.66&  -0.12\\
SE-SIG-DIFF& \bm{\textcolor{red}{2.23}} & \bm{\textcolor{red}{1.06}} & \bm{\textcolor{red}{1.41}} & \bm{\textcolor{red}{2.70}}& \bm{\textcolor{red}{0.74}}\\
\bottomrule 
\addlinespace[2pt]
$Z_{\text{score}}=2, w=90$ & Gruop 1 & Gruop 2 
& Gruop 3 & Gruop 4 &  Gruop 5  \\
\midrule 
SIG& 0.13& -0.08& 0.72& 0.48&  0.57\\
SE-SIG& 0.31& -0.04& 0.91& -0.26&  0.69\\
SE-SIG-DIFF& \bm{\textcolor{red}{1.51}} & \bm{\textcolor{red}{0.62}}& \bm{\textcolor{red}{0.96}}& \bm{\textcolor{red}{1.15}}& \bm{\textcolor{red}{1.48}}\\
\bottomrule 
\addlinespace[2pt]
$Z_{\text{score}}=2.5, w=30$ & Gruop 1 & Gruop 2 
& Gruop 3 & Gruop 4 &  Gruop 5  \\
\midrule 
SIG& 0.64& 0.67& 0.47& 0.26&  0.39\\
SE-SIG& 1.31& 1.29& 1.32& 0.27&  0.29\\
SE-SIG-DIFF& \bm{\textcolor{red}{1.71}} & \bm{\textcolor{red}{1.64}}& \bm{\textcolor{red}{1.97}}& \bm{\textcolor{red}{1.03}}& \bm{\textcolor{red}{1.34}}\\
\bottomrule 
\addlinespace[2pt]
$Z_{\text{score}}=2.5, w=60$ & Gruop 1 & Gruop 2 
& Gruop 3 & Gruop 4 &  Gruop 5  \\
\midrule 
SIG& -1.2& 0.18& -0.29& -0.60&  -0.43\\
SE-SIG& 0.82& 0.72& 0.15& -0.04&  \bm{\textcolor{red}{1.94}}\\
SE-SIG-DIFF& \bm{\textcolor{red}{0.86}} & \bm{\textcolor{red}{1.95}}& \bm{\textcolor{red}{1.07}}& \bm{\textcolor{red}{0.33}}& 1.86\\
\bottomrule 
\addlinespace[2pt]
$Z_{\text{score}}=2.5, w=90$ & Gruop 1 & Gruop 2 
& Gruop 3 & Gruop 4 &  Gruop 5 \\
\midrule 
SIG& -0.93& -0.15& 0.35& -0.31&  0.74\\
SE-SIG& 1.45& \bm{\textcolor{red}{0.63}}& 0.91& 0.12&  0.85\\
SE-SIG-DIFF& \bm{\textcolor{red}{1.52}} & 0.53& \bm{\textcolor{red}{1.44}}& \bm{\textcolor{red}{1.22}}& \bm{\textcolor{red}{1.15}}\\
\bottomrule 
\end{tabular}} \label{robust}
\end{table}

From Table \ref{robust}, we find that no matter how we configure the window size $\omega$ and $Z_{\text{score}}$, the Sharpe ratio is significantly improved, which demonstrates the robustness and general applicability of our method. Then we need to explore whether the improvement of Sharpe ratio is statistically significant. The Sharpe ratio, as a key measure of risk-adjusted performance, has attracted extensive attention regarding its statistical inference. Jobson and Korkie \citep{Jobson1981} were the first to derive a test statistic (JK test) for comparing the Sharpe ratios of two investment portfolios, providing an important foundation for subsequent research. However, their test suffers from notable finite-sample biases. To address this issue, Memmel \citep{Memmel2003} directly corrected the JK test by revising the variance estimation formula to improve its finite-sample properties, resulting in the so-called JK–Memmel test, which improves accuracy in small samples and has since been widely applied in both academia and practice. Applying this method, we use hypothesis testing to verify the efficiency of SE-SIG-DIFF method. First, we define the null hypothesis ($H_0$) and alternative hypothesis ($H_1$):

$$H_0:\Delta Sharpe\leq0,\ H_1:\Delta Sharpe>0,\ \Delta Sharpe= S_{\text{se-sig-diff}} \ -\ S_{\text{no-sig}},$$
where $S_{\text{se-sig-diff}}$ and $S_{\text{no-sig}}$ represent the Sharpe ratio calculated by excess return rate of the SE-SIG-DIFF strategy and the normal pair trading strategy respectively. Then we define the statistic $F$ as:
$$F=\frac{\Delta Sharpe }{\sqrt{\hat{Var}(\Delta Sharpe)}}, \quad p_{\text{value}}=1-\Phi(F),$$
and
$$\hat{Var}(\Delta Sharpe)=\hat{Var}(S_{\text{se-sig-diff}})+\hat{Var}(S_{\text{no-sig}})-\hat{Cov}(S_{\text{se-sig-diff}},S_{\text{no-sig}}),$$ 
where $\hat{Var}$ and $\hat{Cov}$ are the estimates of variance and covariance, $\Phi$ denotes the standard normal cumulative distribution function. For the sake of readability, only the calculation results is represented here, we recommend readers who are interested in the detailed estimate method of $\hat{Var}$ and $\hat{Cov}$ to refer to famous works \citep{Jobson1981} and \citep{Memmel2003}. The significance level $\alpha$ is set as 0.05, if $p_{\text{value}}<\alpha$, then we shall reject $H_0$, which means that our strategy significantly improves the Sharpe ratio. The $p_{\text{value}}$ in the above test in different groups with the conditions $w=60, Z_{\text{score}}=2$ is shown in Table \ref{pvalue}.

\begin{table}[H]
\centering
\caption{$p_{\text{value}}$ of Sharpe Ratio in Hypothesis Testing}
\resizebox{0.7\textwidth}{!}{
\begin{tabular}{lcccccc} 
\toprule 
Group 1 & AUAG & AUAL & AUSN & ALAG & AGSN & ALSN\\
$p_{\text{value}}$ & $0.015$ & $3.18\times 10^{-9}$ & $7.81\times 10^{-13}$ & $3.25\times 10^{-5}$ & $3.53\times 10^{-11}$ & $3.38\times 10^{-12}$\\
\midrule 
Group 2 & CB & CCF & CM & BCF & BM & MCF\\
$p_{\text{value}}$ & $0.0121$ & $7.19\times 10^{-11}$ & $0.0089$ & $2.67\times 10^{-17}$ & $7.96 \times 10^{-5}$ & $9.64 \times 10^{-7}$\\
\midrule 
Group 3 & MASC & MAY & MARB & SCY & SCRB & RBY\\
$p_{\text{value}}$ & $5.11\times 10^{-11}$ & $0.0393$ & $0.0101$ & $3.48 \times 10^{-18}$ & $5.46\times 10^{-7}$ & $3.62\times 10^{-14}$\\
\midrule 
Group 4 & GCSI & GCPA & SIPA & -- & -- & --\\ 
$p_{\text{value}}$ & $4.35\times 10^{-33}$ & $6.32\times 10^{-21}$ & $6.32\times 10^{-15}$ & -- & -- & --\\ 
\midrule 
Group 5 & ZCZW & ZCZS & ZCZL & ZWZS & ZWZL & ZSZL\\
$p_{\text{value}}$ & $6.59\times 10^{-4}$ & $0.0116$ & $0.0211$ & $0.0129$ & $4.05\times 10^{-6}$ & $7.81\times 10^{-7}$\\
\bottomrule 
\end{tabular}
}
\label{pvalue}
\end{table}

Evidently, the $p_{\text{value}}$ for all groups are substantially lower than 0.05, indicating that the improvement achieved by strategy SE-SIG-DIFF is statistically significant. This conclusion remains consistent across different $w$ and $Z_{\text{score}}$, in the interest of brevity, these specific results are not presented here.

\section{Conclusion}
Our study explores an application of the signature method in medium and high-frequency tradings of futures and demonstrates the use of nonlinear features, via data signatures, in arbitrage-based strategies. We decompose the signature into segmented signature and path direction, which have been proven to be more efficient and can significantly help enhance the indicators for price deviation signals in pairs trading. The comprehensive numerical results show that there is an advantage of the SE-SIG-DIFF strategy we proposed in the present work over the traditional signature in the performance. According to the empirical results and statistical test, we believe the the strategy based on segmented signature is profitable, low-risk, stable, interpretable, and controllable strategy, which holds considerable potential for practical application.

The present study contributes to quantitative finance in the following aspects. In the field of trading strategies, we pioneer the use of signatures as filter signals for pair trading, which significantly improve on Sharpe ratio of traditional pair trading strategies. We have discovered segmented signature which has advantage over traditional signature in literature, improving the interpretability of signatures. We demonstrate that segmented signatures enhance existing strategies and indicators, and achieving notably significant results. As far as for data analysis and statistics, we have proposed a more interpretable method for extracting nonlinear features in high-frequency, complex data and have demonstrated their effectiveness. We believe that the present study shall inspire further research and applications of segmented signatures in financial market data analysis and quantitative strategies.

In further research, we attempt to extend our method to various assets such as stocks and options in different markets,including simulation and real trading, to present the efficiency of our strategy in real world deployment. Additionally, simulations under real market conditions could provide deeper insights into its practical performance, including the impact of order execution, liquidity, and evolving market regimes.

\newpage
\section*{Data Availability Statement}

The data that support the findings of this study are available from the corresponding author upon reasonable request.

\section*{Declaration of Interests}

The authors report no conflict of interest.

\bibliographystyle{apacite}
\bibliography{bibliography.bib}

\end{document}